\documentclass[letterpaper, 10 pt, conference]{ieeeconf}  

\IEEEoverridecommandlockouts                              
\usepackage{graphics} 
\usepackage{epsfig} 
\usepackage{mathptmx} 
\usepackage{times} 
\usepackage{amsmath} 
\usepackage{amssymb}  
\usepackage{mathrsfs}
\usepackage{xcolor}
\usepackage{colortbl}
\usepackage{hyperref}
\usepackage{cite}
\usepackage{siunitx}
\usepackage[version=4]{mhchem}
\usepackage{algorithm}
\usepackage{algpseudocode}
\usepackage{booktabs}
\usepackage{tikz}
\usepackage{pgfplots}
\pgfplotsset{compat=newest}
\usepackage{circuitikz}
\usepackage{bm}
\usepackage{subcaption}
\usepackage{stfloats}

\newcommand{\dt}[1]{\frac{\text{d}{#1}}{\text{d}\textit{t}}}

\newcommand{\sectionrefname}{\textit{Section }}
\usepackage{caption}
\title{\LARGE \bf
Distributed Nonlinear Model Predictive Control \\ for District Heating Networks
}
\author{
    Alessandro Bettoni\\
    \small \textit{Politecnico di Milano}\\
    \small \textit{Milano, Italy}\\
    \and
    Giacomo Mastroddi\\
    \small \textit{ETH Zurich}\\
    \small \textit{Zurich, Switzerland}\\
    \and
    Marco Muttoni\\
    \small \textit{ETH Zurich}\\
    \small \textit{Zurich, Switzerland}\\
}

\begin{document}

\maketitle
\thispagestyle{empty}
\pagestyle{empty}

\begin{abstract}
This paper presents a distributed nonlinear model predictive control that uses alternating direction method of multipliers for district heating networks. Exploiting a graph-based modeling of the thermal dynamics, our controller optimizes the mass flow absorption of buildings in a distributed cooperative scheme that mediates between the superior performance of the centralized control and the privacy preservation of the decentralized schemes.
A benchmark three-building network simulation is used to compare the performance of the proposed solution with a decentralized model predictive control scheme. \newline
\end{abstract}


\section{Introduction}
\label{sec:introduction}
In the context of sustainable energy strategies, improvements in the control of District Heating Networks (DHNs) stand out as a key player in the transition to greener and more efficient urban energy systems. In particular, the potential impact of innovations in this field can be understood by considering that buildings account for \SI{40}{\percent} of energy consumption and \SI{36}{\percent} of direct and indirect energy-related greenhouse gas emissions in the EU \cite{eu2023}. Exploring technologies to use all of these resources more effectively is therefore paramount.

To address the complexities of DHN dynamics, a number of different modeling methods have been examined in the literature.  A preliminary approach is represented by \textit{data-driven} techniques that have demonstrated potential in forecasting heat demands and diagnosing network issues \cite{ML_DHModel}. However, these techniques require extensive historical data and do not provide insights into individual building demands. 
Alternatively, \textit{physics-based} methods provide a more profound comprehension of network dynamics, allowing for the capture of complex interactions within the network itself\cite{PB_DHModel}. Nevertheless, these models are often labor-intensive to calibrate and simulate, rendering them less practical for control design purposes. 
A variant suitable for large networks is based on aggregation of agents of the network into substations, enabling a more rapid simulation but losing fine granularity required for controlling individual buildings \cite{SUBST_DHModel}.
Finally, \textit{graph-based} methods exploit the inherent structure of DHNs, using graph representations of the topology to facilitate various modeling tasks \cite{GB_DHModel}, \cite{Model_GB_UsingFlexibility}. \newline Regardless of the model selected to describe district heating, the choice of controller is then critical in order to achieve a satisfactory level of performance with the available computational resources and constraints. In recent decades, a substantial proportion of the literature has concentrated on the study of Model Predictive Control (MPC) due to its intrinsic capability to integrate model knowledge, future anticipation, constraints and the optimization of multiple trades (e.g. comfort, efficiency, \ce{CO2} emissions). More specifically, research regarding MPC of DHNs can be outlined as follows:
\begin{itemize}
    \item \textit{Network-Only Dynamics}: The controller's sole focus is on the thermal and hydraulic dynamics of the DHN, simplifying the control problem and facilitating the understanding of network-specific behaviors and constraints\cite{MPC_onlyNetworks2}, \cite{MPC_onlyNetworks1}.
    
    \item \textit{Network and Building Dynamics}: The modeling is extended to include the buildings connected to the DHN, thereby enhancing the accuracy of the predicted energy demands \cite{MPC_NetworkandBuidling}.
    
    \item \textit{Network and Storage Dynamics}: The inclusion of thermal storage tanks within the DHN enhances more flexibility and cost-effectiveness. For instance, it is possible to incorporate the thermal capacitance of the network and the thermal energy storage tank, resulting in significant savings during peak demand periods \cite{MPC_StorageandNetwork}.
    \item \textit{Network, Building, and Storage Dynamics}: The integration of network, building, and storage models allows for the combination of the advantages discussed in the two preceding points, albeit with increased effort in terms of modeling, parameter identification and computational requirements \cite{CentralizedContr}. It is of importance to highlight the possibility to integrate these models into energy hub optimization to obtain a comprehensive control that also includes heat-electricity conversions\cite{MPC_Distributed_EH}.
\end{itemize}
In light of these variants, a number of different architectural approaches can be distinguished: in \textit{centralised MPC} (CMPC), a single multivariable controller operates according to a comprehensive dynamic model and communicates with all agents \cite{MPC_StorageandNetwork}, \cite{CentralizedContr}. Although CMPC has the potential to achieve a global optimum for the entire DHN, it faces challenges related to fault-tolerance, scalability, and data-gathering. Indeed, the optimization problem is often large and non-convex, making solutions computationally intensive. Furthermore, the central controller must collect comprehensive data on demand characteristics, raising concerns regarding privacy and data-security. Lastly, a fault on the controller can compromise the operation on the entire system, limiting considerably its robustness.
To address these issues, a \textit{decentralized} strategy (DecMPC)  can be employed to design controllers that function autonomously, without direct communication \cite{Bemporad2010}. Privacy is ensured and the scalability limitations are mitigated by the decomposition of the overall problem into simpler and smaller problems that can be solved in parallel. One drawback of this control method is that each agent lacks a comprehensive understanding of the broader network dynamics, which may lead to suboptimal outcomes in terms of global cost optimization.
A solution that mediates between the two previous dichotomous approaches is represented by \textit{distributed MPC} (DMPC). This scheme employ multiple controllers that optimize local cost functions but coordinate their operations through communication, enabling better handling of interactions compared to decentralized control, while also being more scalable and fault-tolerant than centralized control \cite{ReviewDistribControllers}.

The main contribution of this paper is the design of a DMPC for DHNs built upon the \textit{graph-based} method discussed in \cite{GB_DHModel}, \cite{Model_GB_UsingFlexibility}. This design enhances the cooperative control proposed in \cite{MPC_Distributed_EH} by considering the effect of network thermal inertia and duct losses.
Each building minimizes discomfort, network losses and auxiliary energy costs, given temperature and pressure constraints as well as demand forecasts.
Our controller is based on a standard Alternating Direction Method of Multipliers (ADMM), where the MPC problem of each agent is iterated at each time step until the difference of the copies of the shared variables converges within a specified tolerance bound, ensuring that the overall network dynamics is respected. At this point, the first optimal control action is implemented and the procedure is repeated for subsequent initial conditions, according to the principle of receding horizon. \newline
While the advantages of this type of model and the architectural choice of control have been well-discussed in the aforementioned literature of DHNs, to the best of the authors' knowledge, the use of DMPC built on such a graph-based description is currently lacking but worthy of further investigation.

The remainder of this paper is organized as follows. \sectionrefname\ref{sec:modeling} discusses the model used to describe the DHNs. \sectionrefname\ref{sec:controlstrategy} formulates the distributed and decentralized control architectures, specifying the exchange of variables required for cooperation in the computation of the optimal control action. \sectionrefname\ref{sec:results} presents the results of a representative three-building network, comparing the performance of the distributed strategy with that of a fully decentralized control. Lastly, \sectionrefname\ref{sec:conclusion} provides final conclusions and directions for future work.

\section*{Notation}
In this work, the set of nodes $\mathcal{V}$ comprises all buildings in the DHN. For a given building $H$, the set of its direct predecessors is denoted by $\mathcal{V}^-$, and the set of direct successors is denoted by $\mathcal{V}^+$. \newline Within the optimization framework, the prediction horizon is discretized into time steps indexed by $k$, spanning from 0 to the prediction horizon length $K$. A shared decision variable $x^{H,\nu}$ is characterized by the subscript $H$, indicating the building to which the variable pertains, and $\nu$, denoting in which subproblem it is used. \newline 
In the context of the iterative ADMM algorithm, the index $i$ represents the current iteration step, $x(k|i)$ denotes the value of the shared variable at time step $k$ during iteration $i$, and the copies of these decision variables communicated to other users are denoted as $\hat{x}^{H,\nu} \big|_i^k$, with a different notation to highlight the fact that they do not belong to the set of decision variables.

Consequently, the trajectories of the decision variables are defined as follows:
\begin{align*}
    \boldsymbol{x}^{H,\nu} &= \left[ x^{H,\nu}(k=0 \,|\, i_\mathrm{last}), \, x^{H,\nu}(k=1 \,|\, i_\mathrm{last}), \, \ldots \right], \\
    \boldsymbol{u}^{H,\nu} &= \left[ u^{H,\nu}(k=0 \,|\, i_\mathrm{last}), \, u^{H,\nu}(k=1 \,|\, i_\mathrm{last}), \, \ldots \right],
\end{align*}
where $i_\mathrm{last}$ denotes the last iteration, either when reaching the maximum number of iterations $i_\mathrm{max}$ or when the error falls below a specified tolerance $\epsilon$. Similarly, the dual variable associated with the shared variable $x$ is denoted as $\lambda_{H,\nu}^x$, while the step size is written as $\alpha_{H,\nu}^x$ and the damping coefficients are represented by $\delta_{H,\nu}^x$.

\section{Dynamic Modeling of District Heating Networks}
\label{sec:modeling}
\begin{figure}[H]
	\centering
	\tikzset{every picture/.style={line width=0.75pt}} 

\begin{tikzpicture}[x=0.75pt,y=0.75pt,yscale=-1,xscale=1]

\draw [fill=white, fill opacity=1] (108.62,160.56) -- (100.2,160.56) -- (100.2,182.85) -- (68.54,182.85) -- (68.54,160.18) -- (60.87,160.18) -- (84.48,141.57) -- cycle;
\draw [color=blue, draw opacity=1] (223.33,245.67) -- (285.33,245.67);
\draw [color=red, draw opacity=1] (264.67,154.67) -- (300,154.67) -- (300,85.67) node[midway, below, rotate=90, black] {{\footnotesize Bypass}} -- (266.53,85.67);
\draw [color=red, draw opacity=1] (33,125.61) -- (82.36,125.61);
\draw [color=red, draw opacity=1] (225.19,176.62) -- (282.69,176.62);
\draw [color=red, draw opacity=1] (146.86,125.62) -- (202,125.62);
\draw [color=red, draw opacity=1] (86.86,125.62) -- (142.36,125.62);
\draw [color=blue, draw opacity=1] (84.7,195.86) -- (144.7,195.86);
\draw [color=blue, draw opacity=1] (144.7,195.86) -- (179.67,195.86);
\draw [color=blue, draw opacity=1] (35.67,195.86) -- (84.7,195.86);
\draw [color=red, draw opacity=1] (84.61,165.78) -- (84.61,127.98);
\draw [color=blue, draw opacity=1] (84.7,195.86) -- (84.7,174.11);
\draw[black, fill=white] (84.7,195.86) circle (2);
\draw [fill=black, fill opacity=1] (84.61,125.62) -- (87.18,130.33) -- (82.04,130.33) -- cycle;
\draw [fill=black, fill opacity=1] (84.61,125.61) -- (80.11,128.3) -- (80.11,122.92) -- cycle;
\draw [fill=black, fill opacity=1] (84.61,125.62) -- (89.11,122.93) -- (89.11,128.31) -- cycle;

\draw[black, fill=white, fill opacity=1, rounded corners] (16.33,111.87) rectangle (47,208.24);

\draw[color=red, draw opacity=1, fill=red, fill opacity=1] (202,125.62) circle (1.38);
\draw [color=blue, draw opacity=1] (179.67,195.78) -- (179.67,154.67) -- (264.67,154.78);
\draw[color=blue, fill=blue] (179.67,195.78) circle (2);
\draw [color=blue, draw opacity=1] (179.67,195.78) -- (179.67,245.67) -- (223.33,245.67);
\draw [color=red, draw opacity=1] (220.69,176.61) -- (202,176.61) -- (202,125.62);
\draw [color=red, draw opacity=1] (202,125.62) -- (202,85.67) -- (262.03,85.61);
\draw [color=red, draw opacity=1] (285.33,245.67) -- (320,245.67) -- (320,176.67) node[midway, below, rotate=90, black] {{\footnotesize Bypass}} -- (287.19,176.62);
\draw (77.1,165.45) -- (92.21,165.45) -- (92.21,174.47) -- (77.1,174.47) -- cycle;
\draw (77.1,165.45) -- (92.21,174.47);

\draw [fill=white, fill opacity=1] (168.62,160.56) -- (160.2,160.56) -- (160.2,182.85) -- (128.54,182.85) -- (128.54,160.18) -- (120.87,160.18) -- (144.48,141.57) -- cycle;
\draw [color=red, draw opacity=1] (144.61,165.78) -- (144.61,127.98);
\draw [color=blue, draw opacity=1] (144.7,195.86) -- (144.7,174.11);
\draw (137.1,165.45) -- (152.21,165.45) -- (152.21,174.47) -- (137.1,174.47) -- cycle;
\draw (137.1,165.45) -- (152.21,174.47);

\draw [fill=black, fill opacity=1] (144.61,125.62) -- (147.18,130.33) -- (142.04,130.33) -- cycle;
\draw [fill=black, fill opacity=1] (144.61,125.61) -- (140.11,128.3) -- (140.11,122.92) -- cycle;
\draw [fill=black, fill opacity=1] (144.61,125.62) -- (149.11,122.93) -- (149.11,128.31) -- cycle;

\draw[black, fill=white] (144.7,195.86) circle (2);
\draw [fill=white, fill opacity=1] (247.29,210.56) -- (238.87,210.56) -- (238.87,232.85) -- (207.2,232.85) -- (207.2,210.18) -- (199.54,210.18) -- (223.14,191.57) -- cycle;
\draw [color=red, draw opacity=1] (223.28,215.78) -- (223.28,178.98);
\draw [color=blue, draw opacity=1] (223.37,245.67) -- (223.37,224.11);
\draw (215.76,215.45) -- (230.88,215.45) -- (230.88,224.47) -- (215.76,224.47) -- cycle;
\draw (215.76,215.45) -- (230.88,224.47);

\draw[black, fill=white] (223.33,245.67) circle (2);
\draw [fill=black, fill opacity=1] (222.94,176.62) -- (225.51,181.33) -- (220.38,181.33) -- cycle;
\draw [fill=black, fill opacity=1] (222.94,176.61) -- (218.44,179.3) -- (218.44,173.92) -- cycle;
\draw [fill=black, fill opacity=1] (222.94,176.62) -- (227.44,173.93) -- (227.44,179.31) -- cycle;

\draw [fill=white, fill opacity=1] (309.29,210.56) -- (300.87,210.56) -- (300.87,232.85) -- (269.2,232.85) -- (269.2,210.18) -- (261.54,210.18) -- (285.14,191.57) -- cycle;
\draw [color=red, draw opacity=1] (285.28,215.78) -- (285.28,177.98);
\draw [color=blue, draw opacity=1] (285.37,245.67) -- (285.37,224.11);
\draw (277.76,215.45) -- (292.88,215.45) -- (292.88,224.47) -- (277.76,224.47) -- cycle;
\draw (277.76,215.45) -- (292.88,224.47);

\draw[black, fill=white] (285.33,245.78) circle (2);
\draw [fill=black, fill opacity=1] (284.94,176.62) -- (287.51,181.33) -- (282.38,181.33) -- cycle;
\draw [fill=black, fill opacity=1] (284.94,176.61) -- (280.44,179.3) -- (280.44,173.92) -- cycle;
\draw [fill=black, fill opacity=1] (284.94,176.62) -- (289.44,173.93) -- (289.44,179.31) -- cycle;

\draw [fill=white, fill opacity=1] (288.62,119.89) -- (280.2,119.89) -- (280.2,142.18) -- (248.54,142.18) -- (248.54,119.52) -- (240.87,119.52) -- (264.48,100.9) -- cycle;
\draw [color=red, draw opacity=1] (264.61,125.11) -- (264.61,87.31);
\draw [color=blue, draw opacity=1] (264.7,155.19) -- (264.7,133.44);
\draw (257.1,124.79) -- (272.21,124.79) -- (272.21,133.81) -- (257.1,133.81) -- cycle;
\draw (257.1,124.79) -- (272.21,133.81);

\draw[black, fill=white] (264.67,154.78) circle (2);
\draw [fill=black, fill opacity=1] (264.28,85.62) -- (266.84,90.33) -- (261.71,90.33) -- cycle;
\draw [fill=black, fill opacity=1] (264.28,85.61) -- (259.78,88.3) -- (259.78,82.92) -- cycle;
\draw [fill=black, fill opacity=1] (264.28,85.62) -- (268.78,82.93) -- (268.78,88.31) -- cycle;

\node[rotate=90] at (31,160.055) {\footnotesize Heat Producer};

\node[rotate=90] at (215,120) {\footnotesize Split Node};

\end{tikzpicture}
	\caption{A generic DHN consisting of a heat producer, five buildings, and a split node. The components are interconnected through hot water supply pipes, indicated in red color, and cold water return pipes, represented in blue.}
	\label{fig:drawGeneralDHN}
\end{figure}
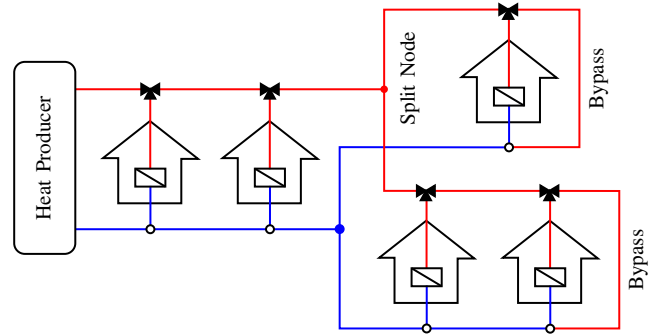

In a DHN, thermal energy produced by a central plant is distributed to several buildings via an extensive system of supply pipes, pumps, distribution valves and splits. As illustrated in \autoref{fig:drawGeneralDHN}, after the heat is exchanged with the buildings, the cooled water is returned to the producer through a separate pipe system. Split nodes are used to bifurcate the flow into segments, forming a complex network that meets geographical constraints and demand characteristics.

\begin{figure}[H]
    \centering
    \begin{minipage}[b]{0.48\linewidth}
        \centering
        \tikzset{every picture/.style={line width=0.75pt}}

\tikzstyle{arrow} = [thick, ->, >=stealth]
\tikzstyle{textBlack} = [black, text centered, text width=1cm, font=\footnotesize]
\tikzstyle{textR} = [blue, above, align=center, pos=0.5, font=\footnotesize]
\tikzstyle{textF} = [red, below, align=center, pos=0.5, font=\footnotesize]
\tikzstyle{textVertical} = [right, align=center, pos=0.5, font=\footnotesize]

\begin{tikzpicture}[x=0.75pt,y=0.75pt,yscale=-1,xscale=1]

\draw [color=blue] (158.59,351.49) -- (158.59,323.31) node[textVertical, color=blue] {S3};

\draw [color=blue, dashed] (158.59,351.49) -- (214.68,351.49) node[textR] {I};

\draw [color=red, dashed] (161.16,243.44) -- (216.67,243.44) node[textF] {O};

\draw [fill=white, fill opacity=1] 
        (158.32,272.76) -- 
        (130.22,295.56) -- 
        (139.35,295.56) -- 
        (139.35,323.31) -- 
        (177.04,323.31) -- 
        (177.04,296.02) -- 
        (187.07,296.02) -- 
        cycle; 

\draw [arrow, color=red] (100.3,243.43) -- (153.13,243.43) node[textF] {F};

\draw [arrow, color=blue] (158.59,351.49) -- (100.22,351.49) node[textR] {R};

\draw [color=red] (158.48,272.76) -- (158.48,249.22) node[textVertical, color=red] {S1};

\draw[black, fill=white] (158.59,351.49) circle (2);
\node[textBlack, below] at (158.59,354.59) {Mixing Node};

\draw [fill=black, fill opacity=1] (158.48,243.44) -- (161.54,249.22) -- (155.43,249.22) -- cycle;

\node[textBlack, above] at (158.48,240.44) {Control Valve};

\draw [fill=black, fill opacity=1] (158.48,243.43) -- (153.13,246.72) -- (153.13,240.14) -- cycle;

\draw [fill=black, fill opacity=1] (158.48,243.44) -- (163.84,240.15) -- (163.84,246.73) -- cycle;

\draw [color=black] (158.48,302.63) node[anchor=south east, font=\footnotesize, color=black] {S2} -- (158.48,272.76);

\draw [fill=white] (149.54,302.01) -- (167.53,302.01) -- (167.53,313.06) -- (149.54,313.06) -- cycle;

\draw (149.54,302.01) -- (167.53,313.06);

\draw [color=black] (158.59,323.31) -- (158.59,312.97);

\end{tikzpicture}
        \subcaption{}
        \label{fig:draw2housesNormal}
    \end{minipage}
    \hfill
    \begin{minipage}[b]{0.48\linewidth}
        \centering
        \tikzset{every picture/.style={line width=0.75pt}}

\tikzstyle{arrow} = [thick, ->, >=stealth]
\tikzstyle{textBlack} = [black, text centered, text width=1cm, font=\footnotesize]
\tikzstyle{textR} = [blue, above, align=center, pos=0.5, font=\footnotesize]
\tikzstyle{textF} = [red, below, align=center, pos=0.5, font=\footnotesize]
\tikzstyle{textVertical} = [right, align=center, pos=0.5, font=\footnotesize]

\begin{tikzpicture}[x=0.75pt,y=0.75pt,yscale=-1,xscale=1]

\draw [color=red] (332.39,350.26) -- (386.89,350.26) -- (386.49,242.01) node[textVertical, left, color=red] {B} -- (334.96,242.22);

\draw [color=blue] (332.39,350.26) -- (332.39,322.09) node[textVertical, color=blue] {S3};

\draw [fill=white, fill opacity=1] 
 (332.12,271.54) -- 
 (304.02,294.34) -- 
 (313.14,294.34) -- 
 (313.14,322.09) -- 
 (350.84,322.09) -- 
 (350.84,294.79) -- 
 (360.86,294.79) -- 
 cycle; 

\draw [arrow, color=red] (274.1,242.2) -- (326.92,242.2) node[textF] {F};

\draw [arrow, color=blue] (332.39,350.26) -- (276.02,350.26) node[textR] {R};

\draw [color=red] (332.28,271.54) -- (332.28,247.99) node[textVertical, color=red] {S1};

\draw[black, fill=white] (332.39,350.26) circle (2);
\node[textBlack, below] at (332.39,353.26) {Mixing Node};

\draw [fill=black, fill opacity=1] (332.28,242.22) -- (335.33,247.99) -- (329.22,247.99) -- cycle;

\node[textBlack, above] at (332.28,239.22) {Control Valve};

\draw [fill=black, fill opacity=1] (332.28,242.2) -- (326.92,245.5) -- (326.92,238.91) -- cycle;

\draw [fill=black, fill opacity=1] (332.28,242.22) -- (337.63,238.93) -- (337.63,245.51) -- cycle;

\draw [color=black] (332.28,301.41) node[anchor=south east, font=\footnotesize, color=black] {S2} -- (332.28,271.54);

\draw [fill=white] (323.33,300.79) -- (341.33,300.79) -- (341.33,311.83) -- (323.33,311.83) -- cycle;

\draw (323.33,300.79) -- (341.33,311.83);

\draw [color=black] (332.39,322.09) -- (332.39,311.75);

\end{tikzpicture}
        \subcaption{}
        \label{fig:draw2housesBypass}
    \end{minipage}
    \caption{Schematics of two buildings and their plumbing system: (a) General user configuration with feeding inlet $F$ and outlet $O$, return inlet $I$ and outlet $R$, hot pipe segment $S1$, heat exchanger $S2$ and cold pipe segment $S3$. (b) User with additional bypass segment $B$.}
    \label{fig:draw2houses}
\end{figure}
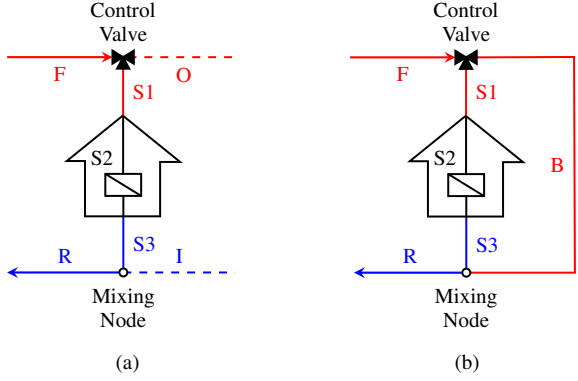

As shown in \autoref{fig:draw2houses}, each building extracts fluids from a supply pipe $F$ using a controlled valve connected to pipe $S1$. The flow then passes through a heat exchanger $S2$, where energy is transferred to the building, and it is conveyed into  return pipe $R$ of the network via a mixing node located at the end of $S3$. Any flow not required by the user is directed to neighboring buildings through the outlet pipe $O$. Similarly, the return flow from the same neighboring building reaches the mixing node via the inlet return $I$. At termination of network branches, buildings are equipped with a bypass $B$ to recirculate unused flow.
As previously mentioned, the notation and graph-based modeling utilized in this paper are grounded in the first-principles approach outlined in \cite{GB_DHModel} and \cite{Model_GB_UsingFlexibility}. To reduce the complexity of the optimization problem and limit the number of shared variables, a \textit{parallel configuration} is selected from the comprehensive treatment provided in the reference study. In this network architecture, which is frequently employed in practice, there is no need for split nodes since hot water is supplied to buildings by a single distribution pipeline, ensuring uniform access to the heating source.

\subsection{Thermal Dynamic Modeling}
The thermal domain employs energy conservation principles to capture the dynamic behavior of the system through the use of lumped parameter temperature models. In particular, the temperature evolution of the pipe segments is described by the following differential equation:
\begin{equation} \label{eq:modelThermalEnergyGeneral}
    \rho c_\mathrm{p} V \dt{T} = \dot{m}_\mathrm{in} c_\mathrm{p} T_\mathrm{in} - \dot{m}_\mathrm{out} c_\mathrm{p} T - \dot{Q}
\end{equation}
where $\rho$ is the density of the fluid used in the DHN (e.g. water), $c_\mathrm{p}$ is its specific heat capacity, $V$ is the volume of the pipe, $T$ denotes the lumped outlet temperature of the mass flow rate $\dot{m}_\mathrm{out}$, $T_\mathrm{in}$ is the inlet temperature associated with rate $\dot{m}_\mathrm{in}$, and $\dot{Q}$ represents the heat loss occurring due to convection in the environment:
\begin{equation} \label{eq:modelThermalHeatLossAmbient}
    \dot{Q} = \dot{Q}_\mathrm{amb} = h A (T - T_\mathrm{amb})
\end{equation}
where $h$ is the heat transfer coefficient and $A$ the surface area of the pipe. \newline
The input energy of the return pipe segment $R$, on the other hand, consists of two streams: the energy from pipe $S3$ and the energy from the return pipe of the successor. This relation is given by
\begin{equation} \label{eq:modelThermalMassReturn}
    \dot{m}_\mathrm{in} c_\mathrm{p} T_\mathrm{in} = \dot{m}_\mathrm{U} c_\mathrm{p} T_\mathrm{S3} + \dot{m}_\mathrm{I} c_\mathrm{p} T_\mathrm{I}.
\end{equation}
In $S2$, heat is extracted by convection and utilized for the regulation of the building temperature. Consequently, the corresponding $\dot{Q}$ to be used in \autoref{eq:modelThermalEnergyGeneral} is slightly different:  
\begin{equation} \label{eq:modelThermalHeatLossBuilding}
    \dot{Q} = h_\mathrm{S2} A_\mathrm{S2} (T_\mathrm{S2} - T_\mathrm{b}).
\end{equation}
Finally, the temperature evolution of a building is influenced by the energy received from the heat exchanger $S2$ and the thermal losses attributed to convective processes at the building surface $A_\mathrm{b}$ with the surrounding ambient:
\begin{equation} \label{eq:modelThermalEnergyBuilding}
    C_\mathrm{b} \dt{T_\mathrm{b}} = h_\mathrm{S2} A_\mathrm{S2} (T_\mathrm{S2} - T_\mathrm{b}) - h_\mathrm{b} A_\mathrm{b} (T_\mathrm{b} - T_\mathrm{amb})
\end{equation}
where the term made by the products $\rho c_\mathrm{p} V$ is grouped as $C_\mathrm{b}$, representing the heat capacity of the building.

\subsection{Fluid Modeling}
In fluid analysis, the network is assumed to maintain an ideal pressure balance, meaning that the sum of all pressure drops occurring around a downstream loop must be zero. This condition can be achieved through the use of throttling valves and booster pumps, ensuring perfect \textit{hydronic balancing} where the demand for mass flow rates in the pipes is met at all times. The relatively rapid fluid dynamics, in comparison to the slower temperature dynamics, can be considered instantaneous and thus described by two algebraic equations: mass balance and pressure drop. 
The absence of storage through each pipe implies that the first fundamental is governed by the law of mass conservation at each node:
\begin{equation}
    \sum \dot{m}_\mathrm{in} = \sum \dot{m}_\mathrm{out}.
\end{equation}
Additionally, in each pipe the fluid experiences a pressure drop $\Delta p$ that is directly proportional to the square of the mass flow rate $\dot{m}$ through the pipe. This relationship is quantified by the Darcy-Weisbach equation:
\begin{equation}
    \Delta p = f \frac{8 L}{\rho \pi^2 D^5} \dot{m}^2
    \label{eq:pressureDarcy}
\end{equation}
where $L$ and $D$ are pipe length and diameter, respectively, and $f$ is Darcy's friction factor, which is assumed to be constant for a complete turbulent fluid motion.

\section{Control Strategy}
\label{sec:controlstrategy}
The following section presents the formulations of DecMPC and DMPC architectures. For both control schemes, it is assumed that the system is fully observable, ensuring that all relevant state variables can be accurately monitored and controlled. This assumption is not critical in the sense that, were it not to hold, one could simply consider the introduction of suitable state observers. In addition, flow delays through the network are not considered in the control action. This simplification is justified due to the high thermal inertia of the network, which results in temperature changes occurring at a much slower rate compared to the water transport delay within the network. Specifically, temperature dynamics range from several hours to days, whereas flow delays are measured in minutes. This substantial difference permits the simplification of the analysis, prioritizing the temperature management over flow timing.

\subsection{Decentralized Model Predictive Control}

\begin{figure*}[t]
    \centering
\tikzstyle{block} = [rectangle, thick, minimum width=1.5cm, minimum height=1.5cm, text centered, text width=1.6cm, draw=black, fill=white, rounded corners]
\tikzstyle{dotBlock} = [rectangle, thick, minimum width=0.5cm, minimum height=1.5cm, text centered, text width=0.5cm, draw=white, fill=white, rounded corners]
\tikzstyle{arrow}        = [black, thick, ->, >=stealth]
\tikzstyle{arrowFeeding} = [red, thick, ->, >=stealth]
\tikzstyle{arrowReturn}  = [blue, thick, ->, >=stealth]
\tikzstyle{textVertical} = [black, right, align=center, pos=0.5]

\begin{tikzpicture}
    \def\up{0.3}
    \def\down{-\up}
    \def\spaceLittle{0.4}
    \def\spaceBig{0.8}
    
    \matrix[
        column 1/.style={column sep=\spaceLittle cm},
        column 2/.style={column sep=\spaceLittle cm},
        column 3/.style={column sep=\spaceBig cm},
        column 4/.style={column sep=\spaceBig cm},
        column 5/.style={column sep=\spaceLittle cm},
        column 6/.style={column sep=\spaceLittle cm},
        row sep=\spaceBig cm]{
        
        \node[block] (HP) {\footnotesize Heat \\ Producer}; &
        \node[dotBlock] (dot1) {$\ldots$}; &
        \node[block] (pred) {\footnotesize DecMPC \\ $\nu^-$}; &
        \node[block] (H) {\footnotesize DecMPC \\ $H$}; &
        \node[block] (succ) {\footnotesize DecMPC \\ $\nu^+$}; &
        \node[dotBlock] (dot2) {$\ldots$}; &
        \node[block] (byp) {\footnotesize DecMPC with Bypass}; &
        \\
        \node[block] (C_HP) {\footnotesize Heat Producer Controller}; &
        &
        \node[dotBlock] (vdot1) {$\vdots$}; &
        \node[block] (C) {\footnotesize Low-Level Valve Controller}; &
        \node[dotBlock] (vdot2) {$\vdots$}; &
        &
        \node[dotBlock] (vdot3) {$\vdots$}; &
        \\
    };
    
        \draw[arrow, <->] (HP) -- (C_HP) node[textVertical] {\footnotesize };
        
        \draw[arrow] (pred) -- (vdot1) node[textVertical] {\footnotesize $\dot{m}_\mathrm{U}^{\nu^-}$};

        \draw[arrow] (H) -- (C) node[textVertical] {\footnotesize $\dot{m}_\mathrm{U}^H$};

        \draw[arrow] (succ) -- (vdot2) node[textVertical] {\footnotesize $\dot{m}_\mathrm{U}^{\nu^+}$};

        \draw[arrow] (byp) -- (vdot3) node[textVertical] {\footnotesize $\ldots$};
    
\end{tikzpicture}
    \caption{Schematic of the DecMPC architecture, showing the heat producer as the starting point and a bypass building as the end point, completing the parallel framework of the network. Each building independently performs nonlinear MPC to optimize its thermal dynamics. The optimal mass flow determined by each building is communicated to a low-level controller, which adjusts the control valve position to follow the mass flow setpoint.}
    \label{fig:DecMPC}
\end{figure*}
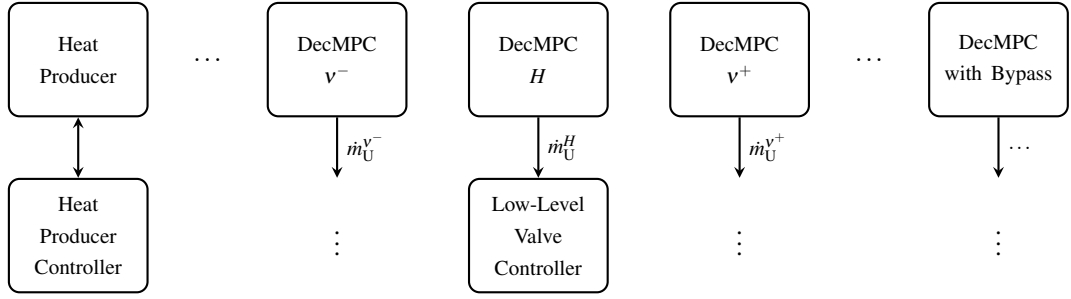

The DecMPC scheme focuses exclusively on the temperature dynamics within a building, specifically $T_\mathrm{S1}$, $T_\mathrm{S2}$, $T_\mathrm{b}$ and $T_\mathrm{S3}$. This approach is designed to operate without the exchange of information between users, as depicted in \autoref{fig:DecMPC}. Each building independently evaluates and determines an optimal flow rate $\dot{m}_\mathrm{U}$, tailored to its thermal needs. The flow rate setpoint is then managed by a low-level controller, not considered in this paper, which adjusts the position of the valve connected to the feeding pipe on the basis of the valve characteristic. \newline The set of decision variables is defined as
\begin{equation}
    \mathcal{D}_\mathrm{private} =
    \left\{
    T_\mathrm{S1}, \,
    T_\mathrm{S2}, \,
    T_\mathrm{S3}, \,
    T_\mathrm{b}, \,
    \dot{m}_\mathrm{U}
    \right\}
\end{equation}
and it is used to solve the following optimization problem:
\begin{equation}        
    \underset{\mathcal{D}_\mathrm{private}}{\min} f_H
\end{equation}
where the cost function $f_H$ associated to building $H$ is:
\begin{equation}
    f_H = \sum_{k=0}^{K} d_H(k) + e_H(k) + c_H(k).
    \label{eq:DecMPC_costfunction}
\end{equation}
The objective is to penalize three factors: temperature discomfort $d_H$ \cite{bhattacharya2018demand}, energy losses of the pipes relative to the ambient temperature $e_H$, and the cost of pumping $c_H$, which is simplified and expressed as a quadratic term of the user mass flow rate \cite{BCANotes}:
\begin{align}
    d_H(k) &= \big\| T_\mathrm{b}(k) - T_\mathrm{b,set}(k) \big\|_{Q_\mathrm{d}}^2 \\
    e_H(k) &= \sum_{i \in \{\text{S1, S3}\}} \big\| T_i(k) - T_\mathrm{amb}(k) \big\|_{Q_\mathrm{loss}^i}^2 \\
    c_H(k) &= \big\| \dot{m}_\mathrm{U}(k) \big\|_{R_\mathrm{cost}}^2.
\end{align}
The optimization is subjected to a reduced nonlinear thermal dynamic as introduced in \autoref{sec:modeling}, which is discretized with a discrete time $T_\mathrm{s}$ using the trapezoidal rule. To find an optimal trade-off among the three cost components, each building must select a flow rate within the constraint of a maximal value $\dot{m}_\mathrm{U}^\mathrm{max}$ calculated by inverting \autoref{eq:pressureDarcy}:
\begin{equation}
    \dot{m} \leq \dot{m}_\mathrm{U}^\mathrm{max} = \sqrt{\frac{\rho \pi^2 D^5}{f 8 L} \Delta p_\mathrm{U}^\mathrm{max}}.
    \label{eq:m_dot_max}
\end{equation}
To ensure sufficient heat supply, the heating plant dynamically adjusts both the temperature and mass flow rate based on the return stream temperature and the bypass mass flow rate. For example, if the return temperature exceeds a specified set value, the plant reduces the feeding temperature. Conversely, if excessive mass flow is discarded in the bypass, the plant responds by decreasing the mass flow output. This reactive adjustment ensures effective thermal regulation across the entire system.

\subsection{Distributed Model Predictive Control}

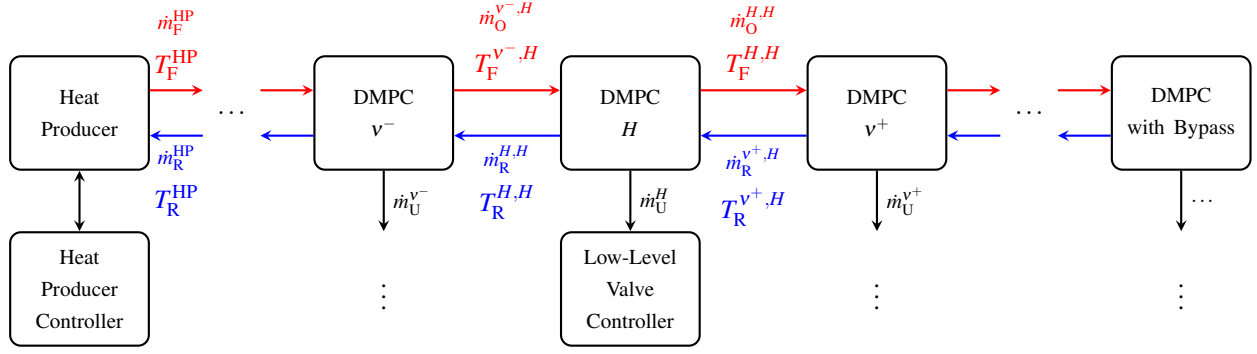
\begin{figure*}[t]
    \centering
\tikzstyle{block} = [rectangle, thick, minimum width=1.5cm, minimum height=1.5cm, text centered, text width=1.6cm, draw=black, fill=white, rounded corners]
\tikzstyle{dotBlock} = [rectangle, thick, minimum width=0.5cm, minimum height=1.5cm, text centered, text width=0.5cm, draw=white, fill=white, rounded corners]
\tikzstyle{arrow}        = [black, thick, ->, >=stealth]
\tikzstyle{arrowFeeding} = [red, thick, ->, >=stealth]
\tikzstyle{arrowReturn}  = [blue, thick, ->, >=stealth]
\tikzstyle{textSent}     = [red,   above, align=center, pos=0.5]
\tikzstyle{textReceived} = [blue,  below, align=center, pos=0.5]
\tikzstyle{textVertical} = [black, right, align=center, pos=0.5]

\begin{tikzpicture}
    \def\up{0.3}
    \def\down{-\up}
    \def\spaceLittle{0.7}
    \def\spaceBig{1.4}
    
    \matrix[
        column 1/.style={column sep=\spaceLittle cm},
        column 2/.style={column sep=\spaceLittle cm},
        column 3/.style={column sep=\spaceBig cm},
        column 4/.style={column sep=\spaceBig cm},
        column 5/.style={column sep=\spaceLittle cm},
        column 6/.style={column sep=\spaceLittle cm},
        row sep=0.8 cm]{
        
        \node[block] (HP) {\footnotesize Heat \\ Producer}; &
        \node[dotBlock] (dot1) {$\ldots$}; &
        \node[block] (pred) {\footnotesize DMPC \\ $\nu^-$}; &
        \node[block] (H) {\footnotesize DMPC \\ $H$}; &
        \node[block] (succ) {\footnotesize DMPC \\ $\nu^+$}; &
        \node[dotBlock] (dot2) {$\ldots$}; &
        \node[block] (byp) {\footnotesize DMPC \\ with Bypass}; &
        \\
        \node[block] (C_HP) {\footnotesize Heat Producer Controller}; &
        &
        \node[dotBlock] (vdot1) {$\vdots$}; &
        \node[block] (C) {\footnotesize Low-Level Valve Controller}; &
        \node[dotBlock] (vdot2) {$\vdots$}; &
        &
        \node[dotBlock] (vdot3) {$\vdots$}; &
        \\
    };
    
        \draw[arrow, <->] (HP) -- (C_HP) node[textVertical] {};
        
        \draw[arrow] (pred) -- (vdot1) node[textVertical] {\footnotesize $\dot{m}_\mathrm{U}^{\nu^-}$};

        \draw[arrow] (H) -- (C) node[textVertical] {\footnotesize $\dot{m}_\mathrm{U}^H$};

        \draw[arrow] (succ) -- (vdot2) node[textVertical] {\footnotesize $\dot{m}_\mathrm{U}^{\nu^+}$};

        \draw[arrow] (byp) -- (vdot3) node[textVertical] {\footnotesize $\ldots$};
        
        \draw[arrowReturn] ([yshift=\down cm]dot1.west) -- ([yshift=\down cm]HP.east) node[textReceived] {\footnotesize $\dot{m}_\mathrm{R}^\mathrm{HP}$ \\[1ex] $T_\mathrm{R}^\mathrm{HP}$};

        \draw[arrowReturn] ([yshift=\down cm]pred.west) -- ([yshift=\down cm]dot1.east) node[textReceived] {};

        \draw[arrowReturn] ([yshift=\down cm]H.west) -- ([yshift=\down cm]pred.east) node[textReceived] {\footnotesize $\dot{m}_\mathrm{R}^{H,H}$ \\[1ex] $T_\mathrm{R}^{H,H}$};

        \draw[arrowReturn] ([yshift=\down cm]succ.west) -- ([yshift=\down cm]H.east) node[textReceived] {\footnotesize $\dot{m}_\mathrm{R}^{\nu^+,H}$ \\[1ex] $T_\mathrm{R}^{\nu^+,H}$};

        \draw[arrowReturn] ([yshift=\down cm]dot2.west) -- ([yshift=\down cm]succ.east) node[textReceived] {};

        \draw[arrowReturn] ([yshift=\down cm]byp.west) -- ([yshift=\down cm]dot2.east) node[textReceived] {};

        \draw[arrowFeeding] ([yshift=\up cm]HP.east) -- ([yshift=\up cm]dot1.west) node[textSent] {\footnotesize $\dot{m}_\mathrm{F}^\mathrm{HP}$ \\[1ex] $T_\mathrm{F}^\mathrm{HP}$};

        \draw[arrowFeeding] ([yshift=\up cm]dot1.east) -- ([yshift=\up cm]pred.west) node[textSent] {};

        \draw[arrowFeeding] ([yshift=\up cm]pred.east) -- ([yshift=\up cm]H.west) node[textSent] {\footnotesize $\dot{m}_\mathrm{O}^{\nu^-,H}$ \\[1ex] $T_\mathrm{F}^{\nu^-,H}$};

        \draw[arrowFeeding] ([yshift=\up cm]H.east) -- ([yshift=\up cm]succ.west) node[textSent] {\footnotesize $\dot{m}_\mathrm{O}^{H,H}$ \\[1ex] $T_\mathrm{F}^{H,H}$};

        \draw[arrowFeeding] ([yshift=\up cm]succ.east) -- ([yshift=\up cm]dot2.west) node[textSent] {};

        \draw[arrowFeeding] ([yshift=\up cm]dot2.east) -- ([yshift=\up cm]byp.west) node[textSent] {};
    
\end{tikzpicture}
    \caption{Schematic of the DMPC architecture, with the heat producer as the starting point and a bypass building as the end point, completing the parallel framework of the network. Each building performs nonlinear MPC based on neighborhood information - feeding in red and return in blue - to optimize its thermal dynamics. The optimal mass flow calculated by each building is given to a low-level controller, which then adjusts the control valve position to follow the mass flow setpoint.}
    \label{fig:DMPC}
\end{figure*}

The DMPC scheme integrates both the feed and return pipes into its control framework, expanding the set of decision variables to include temperatures $T_\mathrm{F}$ and $T_\mathrm{R}$ with associated mass flow rates $\dot{m}_\mathrm{F}$ and $\dot{m}_\mathrm{R}$, as well as the feed inlet temperature $T^\mathcal{\nu-}_\mathrm{F}$ and return inlet temperature $T^\mathcal{\nu+}_\mathrm{R}$. In this comprehensive approach, each building within the DHN implements the nonlinear MPC while considering neighbor conditions and operational constraints. As depicted in \autoref{fig:DMPC}, each building receives information about the feed stream from its predecessor and the return flow from its successor to then solve the following optimization problem:
\begin{equation} \label{eq:DMPC_optProb}
    \underset{\mathcal{D}}{\min} \, g_\mathrm{H}
\end{equation}
where $g_\mathrm{H}$ represents the cost function for building H, containing the same terms as in \autoref{eq:DecMPC_costfunction}. Unlike the DecMPC scheme, $g_\mathrm{H}$ extends the energy efficiency term to penalize losses in the feeding and return streams, as well as in the bypass stream for buildings at the end of a branch: 
\begin{equation}
    e_H(k) = \sum_{i \in \{\text{F, S1, S3, R, (B)}\}} \big\| T_i(k) - T_\mathrm{amb}(k) \big\|_{Q_\mathrm{loss}^i}^2.
\end{equation}
The problem is subjected to the discretized dynamics of the building and the user mass flow rate is constrained by $\dot{m}_\mathrm{U}^\mathrm{max}$ according to \autoref{eq:m_dot_max}, ensuring operational feasibility. Consequently, each building communicates its optimal mass flow to the low-level valve controller, and the first building sends the DHN thermal demand request to the heat producer. For optimal solution finding, the DMPC scheme employs the ADMM algorithm.

\subsubsection*{Alternating Direction Method of Multipliers}

ADMM is an iterative optimization technique that converges to the optimal solution by solving subproblems separately and penalizing deviations in shared variables. In the context of a DHN, the optimization problem naturally decomposes, with shared variables being easily identified. In particular, the set of variables shared with the direct predecessor is
\begin{equation}
    \mathcal{D}^- =
    \left\{
    \dot{m}_\mathrm{O}^{\nu^-,H}, \,
    T_\mathrm{F}^{\nu^-,H}, \,
    \dot{m}_\mathrm{R}^{H,H}, \,
    T_\mathrm{R}^{H,H}
    \right\}
\end{equation}
while the set of variables shared with the direct successor is
\begin{equation}
    \mathcal{D}^+ =
    \left\{
    \dot{m}_\mathrm{O}^{H,H}, \,
    T_\mathrm{F}^{H,H}, \,
    \dot{m}_\mathrm{R}^{\nu^+,H}, \,
    T_\mathrm{R}^{\nu^+,H}
    \right\}
\end{equation}
resulting in the following set of decision variables for the optimization problem defined in \autoref{eq:DMPC_optProb}:
\begin{equation}
    \mathcal{D} =
    \left\{
    \mathcal{D}_\mathrm{private}, \,
    \mathcal{D}^-, \,
    \mathcal{D}^+
    \right\}.
\end{equation}

\begin{algorithm}[!ht]
\caption{ADMM for DMPC}
\label{alg:DMPC}
\begin{algorithmic}[1]
\small
\State \textbf{Offline:}
\State Select $T_\mathrm{s}$, $K$, $T$ and stop conditions $i_\mathrm{max}$ and $\epsilon$
\State Select tuning parameters for all buildings $Q$, $R$, $\delta$, $\alpha$
\State \textbf{Online:}
\State Initialize the states $x$ for all buildings
\For{$t = 0$ to $T$}
    \State Update initial conditions for all buildings' MPC
    \While{$i < i_\mathrm{max}$ and $\text{error} < \epsilon$}
        \For{$v \in \mathcal{V}$}
            \State Solve MPC for building $v$
            \If{$v$ is the first building}
            \State Share trajectories of $\mathcal{D}^+$ with $\mathcal{V}_v^+$
            \ElsIf{$v$ has bypass pipe}
            \State Share trajectories of $\mathcal{D}^-$ with $\mathcal{V}_v^-$
            \Else
            \State \parbox[t]{0.7\linewidth}{Share trajectories of $\mathcal{D}^-$ with $\mathcal{V}_v^-$ and $\mathcal{D}^+$ with $\mathcal{V}_v^+$}
            \EndIf
            \State $v \gets v^+$
        \EndFor
        \State $\text{error} \gets (x - \hat{x})$
        \State Update multipliers: $\lambda \gets \lambda + \alpha \cdot \text{error}$
        \State $i \gets i + 1$
    \EndWhile
    \State Apply $\boldsymbol{u}^*(k=0)$ for all buildings
    \State Advance simulation
    \State $t \gets t + 1$
\EndFor
\end{algorithmic}
\end{algorithm}

As described in Alg.~\ref{alg:DMPC}, building $H$ not only addresses its own thermal demands but also considers the thermal exchange requirements with its neighboring buildings. The objective is to iteratively minimize discrepancies and achieve consensus among the subproblems. To accomplish this, the cost function is defined to balance the individual optimization goals of each building while penalizing the differences between building $H$'s shared thermal variables and the corresponding variables copies provided by neighbors:
\begin{alignat}{2}
    &g_H = \sum_{k=0}^{K} d_H(k) + e_H(k) + c_H(k) && \\
    &+ \sum_{x \in \mathcal{D}^-} \lambda_{H,\nu^-}^x \big|_i^k \cdot \left(x(k|i) - \hat{x} \big|_{i}^k \right) 
    &&+ \frac{\delta_{H,\nu^-}^x}{2} \left\| x(k|i) - \hat{x} \big|_{i}^k \right\|^2 \notag \\
    &+ \sum_{x \in \mathcal{D}^+} \lambda_{H,\nu^+}^x \big|_i^k \cdot \left(x(k|i) - \hat{x} \big|_{i-1}^k \right) 
    &&+ \frac{\delta_{H,\nu^+}^x}{2} \left\| x(k|i) - \hat{x} \big|_{i-1}^k \right\|^2. \notag
\end{alignat}
This cost function includes both linear terms, represented by the dual variables $\lambda_{H,\nu^-}^x$ and $\lambda_{H,\nu^+}^x$, and quadratic augmented components, weighted by the damping coefficients $\delta_{H,\nu^-}^x$ and $\delta_{H,\nu^+}^x$. 
At every iteration, the dual variables of the discrepancies are updated as follows:
\begin{align}
    \boldsymbol{\lambda}_{H,\nu^-}^x \big|_{i+1} &= \boldsymbol{\lambda}_{H,\nu^-}^x \big|_i + \alpha_{H,\nu^-}^x \cdot \left( \boldsymbol{x} \big|_{i} - \boldsymbol{\hat{x}} \big|_{i} \right) \\
    \boldsymbol{\lambda}_{H,\nu^+}^x \big|_{i+1} &= \boldsymbol{\lambda}_{H,\nu^+}^x \big|_i + \alpha_{H,\nu^+}^x \cdot \left( \boldsymbol{x} \big|_{i} - \boldsymbol{\hat{x}} \big|_i \right).
\end{align}
These updates continue according to the step sizes $\alpha_{H,\nu^-}^x$ and $\alpha_{H,\nu^+}^x$ until the differences converge to a predefined tolerance level $\epsilon$, ensuring that the solution has reached consensus across all subproblems. If convergence is not achieved within the maximum number of iterations, the process terminates to prevent indefinite computations. 

\section{Simulation Results}
\label{sec:results}
\subsection{Description of Case Study}
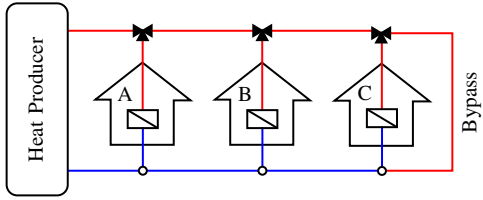
\begin{figure}[H]
	\centering
	\tikzset{every picture/.style={line width=0.75pt}} 

\begin{tikzpicture}[x=0.75pt,y=0.75pt,yscale=-1,xscale=1]

\draw [color=red, draw opacity=1] (33,125.61) -- (82.36,125.61);
\draw [color=red, draw opacity=1] (146.86,125.62) -- (202.36,125.62);
\draw [color=red, draw opacity=1] (86.86,125.62) -- (142.36,125.62);

\draw [color=blue, draw opacity=1] (84.7,195.86) -- (144.7,195.86);
\draw [color=blue, draw opacity=1] (144.7,195.86) -- (204.7,195.86);
\draw [color=blue, draw opacity=1] (35.67,195.86) -- (84.7,195.86);

\draw [fill=white, fill opacity=1] (108.62,160.56) -- (100.2,160.56) -- (100.2,182.85) -- (68.54,182.85) -- (68.54,160.18) -- (60.87,160.18) -- (84.48,141.57) -- cycle;
\draw [color=red, draw opacity=1] (84.61,165.78) node[anchor=south east, font=\footnotesize, color=black] {A} -- (84.61,127.98);
\draw [color=blue, draw opacity=1] (84.7,195.86) -- (84.7,174.11);
\draw (77.1,165.45) -- (92.21,165.45) -- (92.21,174.47) -- (77.1,174.47) -- cycle;
\draw (77.1,165.45) -- (92.21,174.47);
\draw [fill=black, fill opacity=1] (84.61,125.62) -- (87.18,130.33) -- (82.04,130.33) -- cycle;
\draw [fill=black, fill opacity=1] (84.61,125.61) -- (80.11,128.3) -- (80.11,122.92) -- cycle;
\draw [fill=black, fill opacity=1] (84.61,125.62) -- (89.11,122.93) -- (89.11,128.31) -- cycle;
\draw[black, fill=white] (84.7,195.86) circle (2);

\draw [fill=white, fill opacity=1] (168.62,160.56) -- (160.2,160.56) -- (160.2,182.85) -- (128.54,182.85) -- (128.54,160.18) -- (120.87,160.18) -- (144.48,141.57) -- cycle;
\draw [color=red, draw opacity=1] (144.61,165.78) node[anchor=south east, font=\footnotesize, color=black] {B} -- (144.61,127.98);
\draw [color=blue, draw opacity=1] (144.7,195.86) -- (144.7,174.11);
\draw (137.1,165.45) -- (152.21,165.45) -- (152.21,174.47) -- (137.1,174.47) -- cycle;
\draw (137.1,165.45) -- (152.21,174.47);
\draw [fill=black, fill opacity=1] (144.61,125.62) -- (147.18,130.33) -- (142.04,130.33) -- cycle;
\draw [fill=black, fill opacity=1] (144.61,125.61) -- (140.11,128.3) -- (140.11,122.92) -- cycle;
\draw [fill=black, fill opacity=1] (144.61,125.62) -- (149.11,122.93) -- (149.11,128.31) -- cycle;
\draw[black, fill=white] (144.7,195.86) circle (2);

\draw [fill=white, fill opacity=1] (228.65,161.08) -- (220.23,161.08) -- (220.23,183.37) -- (188.57,183.37) -- (188.57,160.71) -- (180.9,160.71) -- (204.51,142.1) -- cycle;
\draw [color=red, draw opacity=1] (204.64,165.3) node[anchor=south east, font=\footnotesize, color=black] {C} -- (204.64,127.5);
\draw [color=blue, draw opacity=1] (204.73,195.38) -- (204.73,173.63);
\draw (197.13,164.98) -- (212.24,164.98) -- (212.24,174) -- (197.13,174) -- cycle;
\draw (197.13,164.98) -- (212.24,174);
\draw [color=red, draw opacity=1] (204.7,195.86) -- (240.03,195.86) -- (240.03,126.86) node[midway, below, rotate=90, black] {{\footnotesize Bypass}} -- (206.56,126.86);
\draw [fill=black, fill opacity=1] (204.31,126.81) -- (206.87,131.52) -- (201.74,131.52) -- cycle;
\draw [fill=black, fill opacity=1] (204.31,126.8) -- (199.81,129.49) -- (199.81,124.11) -- cycle;
\draw [fill=black, fill opacity=1] (204.31,126.81) -- (208.81,124.12) -- (208.81,129.5) -- cycle;
\draw[black, fill=white] (204.7,195.97) circle (2);

\draw[black, fill=white, fill opacity=1, rounded corners] (16.33,111.87) rectangle (47,208.24);
\node[rotate=90] at (31,160.055) {\footnotesize Heat Producer};

\end{tikzpicture}
	\caption{Schematic of the benchmark \textit{parallel configuration} network used in the simulation, consisting of three buildings $A$, $B$, and $C$, a heat producer and a bypass pipe for building $C$.}
	\label{fig:draw3houses}
\end{figure}

The simulation is composed of two distinct parts, designed with the objective of achieving a specific analysis and performed on the three-building network indicated in \autoref{fig:draw3houses}. All the simulation codes used in this work are accessible on our GitHub repository\footnote{\tiny \url{https://github.com/Jackmastro/Distributed_MPC_ATIC.git}}.
The first simulation is intended to evaluate the effectiveness of the DMPC in forecasting and tracking a time-varying setpoint, starting from initial conditions that differ from the reference value.
In contrast, the second simulation compares the control actions of DecMPC and DMPC to confirm the advantage of the latter in terms of fair resource allocation, particularly in the context of scarce heat resources (i.e. saturation of the mass flow of the heat producer HP).
\autoref{table:paramsGeneral} and \autoref{table:paramsPipes} present the values of the parameters used for both simulations, some of which are sourced from \cite{GB_DHModel}, while others are derived from \cite{BCANotes}. In order to streamline the analysis, it is assumed that the three buildings have identical parameters.

\begin{table}[H]
\caption{General parameters.}
\label{table:paramsGeneral}
\centering
\begin{tabular}{l c S[table-format=4.3, table-number-alignment=center] c}
\toprule
    Parameter & Symbol & $\text{Value}$ & Unit \\
\midrule
    Water density & $\rho$ & 971 & $\si{\kilo\gram/\cubic\meter}$ \\[0.5ex]
    Water specific heat capacity & $c_\mathrm{p}$ & 4179 & $\si{\joule/\kilo\gram/\kelvin}$ \\[0.5ex]
    Darcy friction factor & $f$ & 0.025 & $-$ \\[0.5ex]
    Ambient temperature at $t=0$ & $T_\mathrm{amb}$ & -5 & $\si{\celsius}$ \\[0.5ex]
    Building heat capacity & $C_\mathrm{b}$ & 50 & $\si{\mega\joule/\kelvin}$ \\[0.5ex]
    Building surface  & $A_\mathrm{b}$ & 200 & $\si{\meter^2}$ \\[0.5ex]
    Max. diff. pressure of pipe $S$ & $\Delta p_\mathrm{S}^\mathrm{max}$ & 0.4 & $\si{\mega \pascal}$ \\[0.5ex]
    HP max. mass flow rate & $\dot{m}_\mathrm{F}^\mathrm{HP,max}$ & 15 & $\si{\kilo\gram/\second}$  \\[0.5ex]
    HP min. temperature & $T_\mathrm{F}^\mathrm{HP,min}$ & 30 & $\si{\celsius}$  \\[0.5ex]
    HP max. temperature & $T_\mathrm{F}^\mathrm{HP,max}$ & 80 & $\si{\celsius}$  \\
\bottomrule
\end{tabular}
\end{table}

\begin{table}[H]
\caption{Geometric parameters of pipes.}
\label{table:paramsPipes}
\centering
\begin{tabular}{p{1.5cm}
                c
                S[table-format=2.1, table-number-alignment=center]
                S[table-format=1.1, table-number-alignment=center]
                S[table-format=3.0, table-number-alignment=center]
                S[table-format=1.1, table-number-alignment=center]
                S[table-format=2.1, table-number-alignment=center]
                c}
\toprule
    Parameter & Sy. & $F$ & $S\text{1}$ & $S\text{2}$ & $S\text{3}$ & $R$ & Unit\\
\midrule
    Length & $L$ & 40 & 5  & 50 &  5 & 40  & $\si{\meter}$ \\
    Diameter & $D$ & 40  & 10 & 7& 10 & 40 & $\si{\centi\meter}$ \\
    Heat transfer coefficient & $h$ & 1.5  & 1.5 & 100 & 1.5 & 1.5 & $\si{\watt/\square\meter/\kelvin}$ \\
\bottomrule
\end{tabular}
\end{table}

\subsection{Results of Tracking Simulation using DMPC}
The setpoint is chosen to implement the night set-back, under the assumption that the buildings in the network are used for commercial or office purposes. In fact, a reduction in the setpoint during periods when no one is expected to be present (e.g., from 6 PM to 6 AM) is an effective method to limit thermal energy consumption \cite{BCANotes} and it can be exploited to evaluate the balance between tracking accuracy and the convergence of shared variables through an appropriate tuning of parameters. 

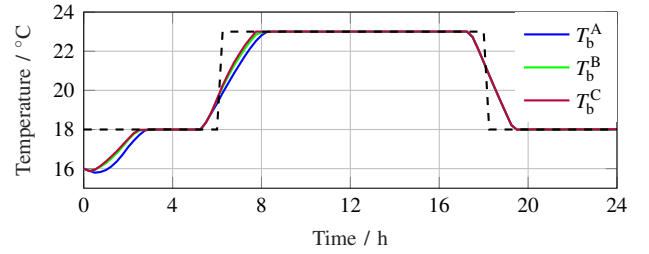
\begin{figure}[H]
	\centering
%
%
\begin{tikzpicture}

\pgfmathsetmacro{\xmin}{0}
\pgfmathsetmacro{\xmax}{1440}
\pgfmathsetmacro{\intervalTicks}{4}
\pgfmathsetmacro{\interval}{60*\intervalTicks}

\begin{axis}[%
width=\linewidth,
height=0.22\textwidth,
unbounded coords=jump,
xmin=\xmin,
xmax=\xmax,
xtick={\xmin,\interval,...,\xmax},
xticklabels={\xmin,\intervalTicks,...,24},
xlabel style={font=\footnotesize\color{white!15!black}},
xlabel={Time / $\si{\hour}$},
ymin=15,
ymax=24,
ylabel style={font=\footnotesize\color{white!15!black}},
ylabel={Temperature / $\si{\celsius}$},
axis background/.style={fill=white},
tick label style={font=\footnotesize},
grid=both,
legend style={at={(1,1)}, anchor=north east, legend cell align=left, draw=none, font=\footnotesize},
]

\addplot [color=blue, thick]
  table[row sep=crcr]{%
0	16\\
15	15.8891848203762\\
30	15.7935268187906\\
45	15.8278524055303\\
60	15.9281346723324\\
75	16.1172950396417\\
90	16.3814683190951\\
105	16.7174483922491\\
120	17.1009107048951\\
135	17.4275438995949\\
150	17.7167774398862\\
165	17.9493727452008\\
180	18.003117990429\\
195	17.9974697628323\\
210	18.0007237014842\\
225	17.9972444463283\\
240	18.0000709495283\\
255	17.9938399807664\\
270	17.9993678637881\\
285	17.9951607601321\\
300	17.9988969015087\\
315	17.9951173734519\\
330	18.3677764762975\\
345	18.8493416115916\\
360	19.3335626418461\\
375	19.8673338527237\\
390	20.3556033489837\\
405	20.845037154674\\
420	21.2989232286577\\
435	21.7485216879916\\
450	22.1598558391333\\
465	22.5219041361054\\
480	22.8108239238912\\
495	22.9939430859083\\
510	23.0014959941234\\
525	22.9990436124039\\
540	22.9996319476276\\
555	22.998802550792\\
570	22.9993735852313\\
585	22.9987161412845\\
600	22.9990777665971\\
615	22.9985424654975\\
630	22.9987875337952\\
645	22.9984600642125\\
660	22.9987390964262\\
675	22.9985428116483\\
690	22.9987519788629\\
705	22.9976993605032\\
720	22.9986437508933\\
735	22.9982652609132\\
750	22.9986719317478\\
765	22.9986700523269\\
780	22.9987410589939\\
795	22.9987162205512\\
810	22.9988022536479\\
825	22.9987771580367\\
840	22.9988631012397\\
855	22.9988472424462\\
870	22.9989695927925\\
885	22.9989282512907\\
900	22.9990082974698\\
915	22.9990156173463\\
930	22.9990929704798\\
945	22.9991106834316\\
960	22.9991933628661\\
975	22.9992169167199\\
990	22.9993012529963\\
1005	22.9993384448144\\
1020	22.999426588655\\
1035	22.999467023908\\
1050	22.7212335805215\\
1065	22.1104140286341\\
1080	21.4406549060217\\
1095	20.7705697263744\\
1110	20.1122184390356\\
1125	19.4676050295495\\
1140	18.8365833480281\\
1155	18.2185690001763\\
1170	17.9917493158162\\
1185	17.9994028085736\\
1200	17.9890470030408\\
1215	18.0017048834526\\
1230	17.9909257729394\\
1245	18.0000029391894\\
1260	17.9950280147402\\
1275	17.9996008517296\\
1290	17.9959672442648\\
1305	17.9995441261612\\
1320	17.9983440468554\\
1335	17.9996830077602\\
1350	17.9991370741762\\
1365	17.9997797267865\\
1380	17.9991778048076\\
1395	17.9997666312352\\
1410	17.9992073968064\\
1425	17.9997107465139\\
1440	17.9979451452191\\
};
\addlegendentry{$T_\mathrm{b}^\mathrm{A}$}

\addplot [color=green, thick]
  table[row sep=crcr]{%
0	16\\
15	15.8799401177343\\
30	15.9323926849459\\
45	16.0961176934541\\
60	16.2778799691622\\
75	16.5252493537893\\
90	16.8015358773056\\
105	17.1265690528164\\
120	17.4662624161081\\
135	17.7627720311041\\
150	17.9413284972139\\
165	17.9881207602369\\
180	17.9974277482128\\
195	18.001801465899\\
210	17.9971733377574\\
225	18.009237958521\\
240	17.9997907483541\\
255	18.0011488366959\\
270	18.000443460489\\
285	18.0008956183592\\
300	18.0016758548119\\
315	18.0022869401139\\
330	18.3624928252045\\
345	18.8986841575257\\
360	19.5211998903975\\
375	20.1648369785223\\
390	20.7149319463001\\
405	21.2366466634847\\
420	21.6897845771585\\
435	22.1226182411255\\
450	22.5050448982374\\
465	22.8425551825778\\
480	22.9587148145349\\
495	22.9862643537824\\
510	23.0016917474861\\
525	22.9983879131519\\
540	22.9948391307428\\
555	22.9955637729134\\
570	22.9960924969201\\
585	22.9981768555035\\
600	23.0001800611836\\
615	23.0015303797078\\
630	23.0010257748152\\
645	23.000439504407\\
660	23.0001536148927\\
675	23.0001020001378\\
690	23.000663055948\\
705	23.0013092479892\\
720	23.0017794888062\\
735	23.0020653583445\\
750	23.0021728667185\\
765	23.0022475082045\\
780	23.002261736937\\
795	23.002248619593\\
810	23.0022102239454\\
825	23.0021520853269\\
840	23.0020935187019\\
855	23.0020432869987\\
870	23.0020041602677\\
885	23.0019744990368\\
900	23.0019352937711\\
915	23.0019221286922\\
930	23.0019135656222\\
945	23.0018859735989\\
960	23.0018788439273\\
975	23.0018897410532\\
990	23.0019160542271\\
1005	23.0017142795403\\
1020	23.0020993028456\\
1035	23.001887962055\\
1050	22.7222525196412\\
1065	22.1111120318383\\
1080	21.4412667679459\\
1095	20.7711468900375\\
1110	20.1127726070749\\
1125	19.4681392760367\\
1140	18.8370988590311\\
1155	18.219063441556\\
1170	17.9915710602509\\
1185	18.0117621536821\\
1200	17.9932796640229\\
1215	18.0065912028502\\
1230	17.9953283874561\\
1245	18.0017629764393\\
1260	17.9977935267853\\
1275	18.0002200464655\\
1290	17.9999711029496\\
1305	18.0013878194709\\
1320	18.0013152183874\\
1335	18.0016464292401\\
1350	18.0013956271836\\
1365	18.0010712988002\\
1380	18.0008512994041\\
1395	18.0007530413551\\
1410	18.0007805912886\\
1425	18.0023997457755\\
1440	18.0035269904266\\
};
\addlegendentry{$T_\mathrm{b}^\mathrm{B}$}

\addplot [color=purple, thick]
  table[row sep=crcr]{%
0	16\\
15	15.8753322133078\\
30	15.9465591099748\\
45	16.1395113558836\\
60	16.3693986411257\\
75	16.6315196416333\\
90	16.9069580132855\\
105	17.2146663029745\\
120	17.5176352788401\\
135	17.8143660267389\\
150	17.9874812380705\\
165	17.9816582755565\\
180	17.9888224969039\\
195	18.0023620375292\\
210	17.9947384721876\\
225	18.0171175276644\\
240	17.9959954383023\\
255	18.0011484365318\\
270	17.9967080625602\\
285	17.997514363017\\
300	17.9981189287432\\
315	17.9972799668687\\
330	18.3497151775807\\
345	18.8902143391707\\
360	19.5635881709656\\
375	20.2501209559658\\
390	20.8433980701143\\
405	21.3942634720888\\
420	21.8577984933059\\
435	22.298121911405\\
450	22.6608312263169\\
465	23.0009313775922\\
480	23.0006924920289\\
495	22.9790026724264\\
510	22.9990506185664\\
525	22.9948997800935\\
540	23.0136256043547\\
555	22.9950839817976\\
570	23.0012195885883\\
585	22.9983728664445\\
600	23.0014948338323\\
615	23.0012851106041\\
630	23.0006961106045\\
645	22.9955655121403\\
660	22.993949040466\\
675	22.9960141978974\\
690	22.9985779202485\\
705	22.9994895158134\\
720	23.0002849873808\\
735	23.0004448832154\\
750	23.0008436234206\\
765	23.0008377215392\\
780	23.0010596642356\\
795	23.0009925071257\\
810	23.0011374192663\\
825	23.0011003491934\\
840	23.0012439805481\\
855	23.0012044183385\\
870	23.0013505022602\\
885	23.0013787292527\\
900	23.0015646623326\\
915	23.0016626039841\\
930	23.0018908487162\\
945	23.0020288731749\\
960	23.0022763626605\\
975	23.0024618166587\\
990	23.0027673102\\
1005	23.0044541068273\\
1020	23.0044411393782\\
1035	23.0046482089428\\
1050	22.7284810913951\\
1065	22.1178959373767\\
1080	21.447981777975\\
1095	20.7776643619132\\
1110	20.1190710298903\\
1125	19.4742201259602\\
1140	18.8429683783946\\
1155	18.2247317921115\\
1170	17.9887688801521\\
1185	18.0082442422632\\
1200	17.9972685853955\\
1215	18.0085498718587\\
1230	17.9968692972405\\
1245	18.0033380680603\\
1260	17.9984002814955\\
1275	18.0011111593529\\
1290	18.0021896338113\\
1305	18.0051105271639\\
1320	18.006477451685\\
1335	18.0069787268507\\
1350	18.0061473534666\\
1365	18.0057092785289\\
1380	18.0052680764679\\
1395	18.0051217568586\\
1410	18.0062309575719\\
1425	18.011802150423\\
1440	18.0149532591717\\
};
\addlegendentry{$T_\mathrm{b}^\mathrm{C}$}

\addplot [color=black, dashed, thick]
  table[row sep=crcr]{%
0	18\\
15	18\\
30	18\\
45	18\\
60	18\\
75	18\\
90	18\\
105	18\\
120	18\\
135	18\\
150	18\\
165	18\\
180	18\\
195	18\\
210	18\\
225	18\\
240	18\\
255	18\\
270	18\\
285	18\\
300	18\\
315	18\\
330	18\\
345	18\\
360	18\\
375	23\\
390	23\\
405	23\\
420	23\\
435	23\\
450	23\\
465	23\\
480	23\\
495	23\\
510	23\\
525	23\\
540	23\\
555	23\\
570	23\\
585	23\\
600	23\\
615	23\\
630	23\\
645	23\\
660	23\\
675	23\\
690	23\\
705	23\\
720	23\\
735	23\\
750	23\\
765	23\\
780	23\\
795	23\\
810	23\\
825	23\\
840	23\\
855	23\\
870	23\\
885	23\\
900	23\\
915	23\\
930	23\\
945	23\\
960	23\\
975	23\\
990	23\\
1005	23\\
1020	23\\
1035	23\\
1050	23\\
1065	23\\
1080	23\\
1095	18\\
1110	18\\
1125	18\\
1140	18\\
1155	18\\
1170	18\\
1185	18\\
1200	18\\
1215	18\\
1230	18\\
1245	18\\
1260	18\\
1275	18\\
1290	18\\
1305	18\\
1320	18\\
1335	18\\
1350	18\\
1365	18\\
1380	18\\
1395	18\\
1410	18\\
1425	18\\
1440	18\\
};

\end{axis}

\end{tikzpicture}%
	\caption{Tracking results for building temperatures using DMPC over a day-long simulation.}
	\label{fig:resultsDMPCTracking24h}
\end{figure}

The tracking reported in \autoref{fig:resultsDMPCTracking24h} is highly satisfactory, even with a variable ambient temperature assumed to be the first half-wave of a sine function, ranging from \SI{-5}{\celsius} at midnight to \SI{5}{\celsius} at noon. This precise tracking can be attributed, besides proper tuning, to the assumptions of perfect model knowledge and exact prediction of ambient temperatures.
The advantage of using MPC is also evident, as it allows for the anticipation of heat demand, enabling heating before 6 AM and limiting heat absorption slightly before 6 PM. In this context, the rapid descent is attributed to the high losses of the buildings to the environment, as no cooling is taken into account.

\begin{figure}[ht]
	\centering
	\input{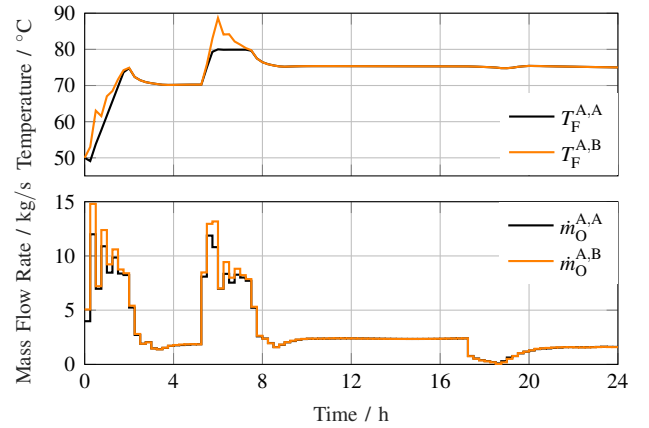}
	\caption{Evolution of the discrepancy for the temperature and mass flow of the feeding outlet of building $A$ over a day-long simulation.}
	\label{fig:resultsDMPCconvergence}
\end{figure}

\autoref{fig:resultsDMPCconvergence} illustrates the evolution of the mismatch between shared and actual temperature and mass flow variables across successive time steps of the controller.
Notwithstanding the arbitrary initial conditions at time $t=0$, selected by the authors to be different from steady-state values in order to assess the resilience of ADMM in achieving convergence, the results reveal that controllers do eventually converge on the shared values quite well. The price to pay for a bad initialization is the considerable number of iterations required to converge. Subsequent iterations, which begin with warm initialization, exhibit instead faster convergence rates and smaller error. 
Finally, the mismatch observed in the temperature variables around the sixth hour can be attributed to the setpoint shift, which causes an abrupt increase in the mass flow rate requested by the users, leading to inaccuracies in both temperatures and mass flows.

\subsection{Results of Comparison Simulation}
The second part of the simulation assumes the initial conditions for each building to be different, with buildings $A$, $B$, and $C$ starting at \SI{15}{\celsius}, \SI{16}{\celsius}, and \SI{17}{\celsius}, respectively. The building closest to the heat producer is selected as the one with the greatest need for energy while the furthest building is considered to have the least immediate need for energy. In addition, a constant ambient temperature of \SI{-5}{\celsius} and a fixed setpoint of \SI{18}{\celsius} are imposed for simplicity.  As detailed in \autoref{table:paramsGeneral}, the maximum mass flow rate in each edifice is constrained by the maximum differential pressure that the pipes can withstand, achieved at approximately \SI{8}{\kilo\gram/\second} in our case. 
The heat producer has a maximum output flow of \SI{15}{\kilo\gram/\second} and therefore it cannot simultaneously meet the maximum demands of all three buildings. This assumption is reasonable given that a DHN cannot be oversized to meet peak demand for cost reasons.  

\autoref{table:paramsDMPC} and \autoref{table:paramsDecMPC} present the parameters of DMPC and DecMPC, respectively, used in the simulation.

\begin{table}[htbp]
\caption{Parameters of the DMPC.}
\label{table:paramsDMPC}
\centering
\begin{tabular}{l c S[table-format=2.2, table-number-alignment=center] c}
\toprule
    Parameter & Symbol & $\text{Value}$ & Unit \\
\midrule
\arrayrulecolor{gray!10}
    Sampling time & $T_\mathrm{s}$ & 15 & $\si{\minute}$ \\
    Prediction horizon length & $K$ & 1 & \si{\hour} \\
    Simulation time & $T$ & {10 - 24} & $\si{\hour}$ \\
    Max. iteration & $i_\mathrm{max}$ & 60 & $-$ \\
    Tolerance of error norm & $\epsilon$ & 1 & $-$ \\
\midrule
    Discomfort weight & $Q_\mathrm{d} $ & 2 & \SI{e5}{} \\[0.2ex]
    Pipe segment $S$ efficiency weight & $Q_\mathrm{loss}^\mathrm{S} $ & {0.01 - 0.8} & $-$ \\[0.2ex]
    Bypass efficiency weight & $Q_\mathrm{loss}^\mathrm{B} $ & 3 & \SI{e4}{} \\
    Pumping weight & $R_\mathrm{cost} $ & 20 & $-$ \\
\midrule
    Damping coefficient for temperatures & $\delta^\mathrm{T}$ & 1.5 & \SI{e5}{} \\
    Damping coefficient for mass flows & $\delta^\mathrm{\dot{m}}$ & 6 & \SI{e5}{} \\
    Step size for temperatures & $\alpha^\mathrm{T}$ & 0.05 & $-$ \\
    Step size for mass flows & $\alpha^\mathrm{\dot{m}}$ & 0.2 & $-$ \\
\arrayrulecolor{black}
\bottomrule
\end{tabular}
\end{table}

\begin{table}[htbp]
\caption{Parameters of the DecMPC.}
\label{table:paramsDecMPC}
\centering
\begin{tabular}{l c S[table-format=2.2, table-number-alignment=center] c}
\toprule
    Parameter & Symbol & $\text{Value}$ & Unit \\
\midrule
\arrayrulecolor{gray!10}
    Sampling time & $T_\mathrm{s}$ & 15 & $\si{\minute}$ \\
    Prediction horizon length& $K$ & 1 & $\si{\hour}$ \\
\midrule
    Discomfort weight & $Q_\mathrm{d} $ & 11 & $-$ \\
    Efficiency weights & $Q_\mathrm{loss} $ & 1.5 & \SI{e-3}{} \\
    Pumping weight & $R_\mathrm{cost} $ & 0.03 & $-$ \\
\arrayrulecolor{black}
\bottomrule
\end{tabular}
\end{table}

These values have been selected to ensure that the efficiency $\eta$ of both control strategies are comparable:
\begin{equation}
    \eta = \frac{\dot{Q}_\mathrm{b}}{\dot{Q}_\mathrm{tot}^\mathrm{HP}}
\end{equation}
where the total power fed in the DHN by the heat producer $\dot{Q}_\mathrm{tot}^\mathrm{HP}$ and the total power absorbed by the buildings $\dot{Q}_\mathrm{b}$ are defined as follows:
\begin{align}
    \dot{Q}_\mathrm{tot}^\mathrm{HP} &= \dot{m}_\mathrm{F}^\mathrm{HP} c_\mathrm{p} T_\mathrm{F}^\mathrm{HP} - \dot{m}_\mathrm{R}^\mathrm{HP} c_\mathrm{p} T_\mathrm{R}^\mathrm{HP} \\
    \dot{Q}_\mathrm{b} &= \sum_{\nu \in \mathcal{V}} h_\mathrm{S2}^\nu A_\mathrm{S2}^\nu \cdot \left( T_\mathrm{S2}^\nu - T_\mathrm{b}^\nu\right).
\end{align}
The comparability between efficiencies, that in the case of our simulation are \SI{60}{\percent} for DMPC and \SI{61}{\percent} for DecMPC, is essential for correctly evaluating the performance of the control strategies. In fact, this prevents any bias in fairness from arising if one control architecture were set to prioritize efficiency to a greater extent than the other.

\begin{figure}[H]
	\centering
%
%
\begin{tikzpicture}

\pgfmathsetmacro{\xmin}{0}
\pgfmathsetmacro{\xmax}{600}
\pgfmathsetmacro{\intervalTicks}{2}
\pgfmathsetmacro{\interval}{60*\intervalTicks}
\pgfmathsetmacro{\height}{0.25\textwidth}

\begin{axis}[%
name=temperaturePlot,
width=\linewidth,
height=\height,
unbounded coords=jump,
xmin=\xmin,
xmax=\xmax,
xtick={\xmin,\interval,...,\xmax},
xticklabels=\empty,
ymin=14,
ymax=19,
ylabel style={font=\footnotesize\color{white!15!black}},
ylabel={Temperature / $\si{\celsius}$},
axis background/.style={fill=white},
tick label style={font=\footnotesize},
grid=both,
legend style={at={(1,0)}, anchor=south east, legend cell align=left, draw=none, font=\footnotesize},
]

\addplot [color=blue, thick]
  table[row sep=crcr]{%
0	15\\
15	15.2546979655494\\
30	15.6393054261881\\
45	15.966146246865\\
60	16.239531872017\\
75	16.484140200362\\
90	16.7158756235906\\
105	16.9386409510151\\
120	17.1480416261134\\
135	17.3374606819893\\
150	17.5025710507744\\
165	17.6410315471181\\
180	17.749617063043\\
195	17.8242549181205\\
210	17.8671614540267\\
225	17.8875858360179\\
240	17.8951220604158\\
255	17.8964075948357\\
270	17.8952849763501\\
285	17.8937196338715\\
300	17.8925482731457\\
315	17.8919708420244\\
330	17.8918857687141\\
345	17.8920972350977\\
360	17.8924333350006\\
375	17.89277499735\\
390	17.8930582127885\\
405	17.8932589863445\\
420	17.8933777938205\\
435	17.8934278428022\\
450	17.8934275959458\\
465	17.8933961176115\\
480	17.8933499784074\\
495	17.8933016434193\\
510	17.8932591914393\\
525	17.8932268179025\\
540	17.8932056665516\\
555	17.8931947850563\\
570	17.8931920214532\\
585	17.8931947888271\\
600	17.8932006137472\\
};
\addlegendentry{$T_\mathrm{b}^\mathrm{A}$}

\addplot [color=green, thick]
  table[row sep=crcr]{%
0	16\\
15	16.0018238742019\\
30	16.2278964432475\\
45	16.5044615267674\\
60	16.7526121724566\\
75	16.9662323288192\\
90	17.1596531200693\\
105	17.3420584659495\\
120	17.5111067521555\\
135	17.6564501830853\\
150	17.7667643050194\\
165	17.8377945913593\\
180	17.8760871280365\\
195	17.8928982220606\\
210	17.897935486691\\
225	17.8975893799774\\
240	17.8954102028607\\
255	17.8930680909696\\
270	17.8911911511447\\
285	17.8899206465469\\
300	17.8892216809102\\
315	17.8890045132862\\
330	17.8891499840515\\
345	17.8895245780036\\
360	17.8900033177899\\
375	17.8904882903249\\
390	17.8909154337747\\
405	17.8912516730572\\
420	17.891487644164\\
435	17.8916299294028\\
450	17.8916944498496\\
465	17.8917012960126\\
480	17.89167089214\\
495	17.8916213492306\\
510	17.8915669057331\\
525	17.8915173493508\\
540	17.8914782050975\\
555	17.8914514584885\\
570	17.8914365371554\\
585	17.8914313129142\\
600	17.8914329862549\\
};
\addlegendentry{$T_\mathrm{b}^\mathrm{B}$}

\addplot [color=purple, thick]
  table[row sep=crcr]{%
0	17\\
15	16.5867645350723\\
30	15.9886716264106\\
45	15.7085682054114\\
60	15.6413622933301\\
75	15.7043660478536\\
90	15.8593849043062\\
105	16.0697197230706\\
120	16.3079509448909\\
135	16.5491680862635\\
150	16.77594259987\\
165	16.982911867439\\
180	17.1697545499468\\
195	17.3380017424327\\
210	17.4888084590083\\
225	17.6209297430345\\
240	17.7293413570347\\
255	17.807215861858\\
270	17.8540230047406\\
285	17.8775803781224\\
300	17.8873006647145\\
315	17.8901022658581\\
330	17.8901204608945\\
345	17.8894625222196\\
360	17.8889816723516\\
375	17.8888815502342\\
390	17.8890847746856\\
405	17.8894417476307\\
420	17.8898233618163\\
435	17.8901484127638\\
450	17.8903807516346\\
465	17.8905163917478\\
480	17.8905699465616\\
495	17.8905639970924\\
510	17.8905219428271\\
525	17.8904638772588\\
540	17.8904047283793\\
555	17.890353945795\\
570	17.8903161456516\\
585	17.8902922442524\\
600	17.8902806975167\\
};
\addlegendentry{$T_\mathrm{b}^\mathrm{C}$}

\addplot [color=black, dashed, thick]
  table[row sep=crcr]{%
0	18\\
600	18\\
};

\end{axis}

\begin{axis}[%
at={(temperaturePlot.below south)},
anchor=north,
yshift=-0.1cm,
width=\linewidth,
height=\height,
unbounded coords=jump,
xmin=\xmin,
xmax=\xmax,
xtick={\xmin,\interval,...,\xmax},
xticklabels={\xmin,\intervalTicks,...,24},
xlabel style={font=\footnotesize\color{white!15!black}},
xlabel={Time / $\si{\hour}$},
ymin=0,
ymax=9,
ylabel style={font=\footnotesize\color{white!15!black}},
ylabel={Mass Flow Rate / $\si{\kilo\gram/\second}$},
axis background/.style={fill=white},
tick label style={font=\footnotesize},
grid=both,
legend style={at={(1,1)}, anchor=north east, legend cell align=left, draw=none, font=\footnotesize},
]

\addplot [color=black]
  table[row sep=crcr]{%
0	8.02665475463358\\
600	8.02665475463358\\
};
\addlegendentry{$\dot{m}_\mathrm{U}^\mathrm{max}$}

\addplot [const plot, color=blue, thick]
  table[row sep=crcr]{%
0	8.02665475463358\\
15	8.02665475463358\\
30	7.71666084227709\\
45	7.20038703553995\\
60	6.65503443969597\\
75	6.03903574926582\\
90	5.47971602796922\\
105	4.98113159265708\\
120	4.4568455207974\\
135	3.87266515594071\\
150	3.20690237332159\\
165	2.48308749173903\\
180	1.84582602589618\\
195	1.43965867227593\\
210	1.23061884208272\\
225	1.1280394115101\\
240	1.07705029104393\\
255	1.05130436499018\\
270	1.03796831449493\\
285	1.03086335128331\\
300	1.02689026476714\\
315	1.02450241018693\\
330	1.02297658949041\\
345	1.02209958269176\\
360	1.02173845932784\\
375	1.02180436764757\\
390	1.02216782345694\\
405	1.02268575760543\\
420	1.02323014903151\\
435	1.02371066423045\\
450	1.0240789005645\\
465	1.02432002683013\\
480	1.02444335544514\\
495	1.0244732048821\\
510	1.02443973654837\\
525	1.0243716925046\\
540	1.02429240744262\\
555	1.02421810467303\\
570	1.02415821750432\\
585	1.0241164506229\\
600	1.02409243879092\\
};
\addlegendentry{$\dot{m}_\mathrm{U}^\mathrm{A}$}

\addplot [const plot, color=green, thick]
  table[row sep=crcr]{%
0	6.97334524536642\\
15	6.90107618874027\\
30	6.39706608023718\\
45	5.9038999976176\\
60	5.42850211917026\\
75	4.96114415792424\\
90	4.43739755441324\\
105	3.79410333902388\\
120	3.02753579718586\\
135	2.27465525652248\\
150	1.72218832117543\\
165	1.40914118702184\\
180	1.24944020366332\\
195	1.16827698739069\\
210	1.12637639576188\\
225	1.10413287475484\\
240	1.09133615217584\\
255	1.08240231959524\\
270	1.07455103807568\\
285	1.06724541179613\\
300	1.06089242628325\\
315	1.05578965517768\\
330	1.05195874519908\\
345	1.04928243147318\\
360	1.04759715781444\\
375	1.046717679383\\
390	1.04644377433346\\
405	1.04657395342199\\
420	1.04692630955052\\
435	1.04735591738007\\
450	1.04776269110629\\
465	1.04808984908937\\
480	1.04831588758971\\
495	1.0484439118401\\
510	1.0484913699761\\
525	1.04848152412672\\
540	1.04843761100684\\
555	1.04837926491594\\
570	1.04832083340978\\
585	1.04827128407733\\
600	1.04823487004076\\
};
\addlegendentry{$\dot{m}_\mathrm{U}^\mathrm{B}$}

\addplot [const plot, color=purple, thick]
  table[row sep=crcr]{%
0	-1.58758257175729e-28\\
15	0.0722690566275078\\
30	0.886273077485739\\
45	1.89571296684245\\
60	2.91646344113376\\
75	3.99982009280993\\
90	5.08288641761754\\
105	6.22476506831904\\
120	6.45163025458873\\
135	5.86666538810867\\
150	5.37946302630668\\
165	4.93348007218217\\
180	4.47148833818167\\
195	3.95110189360858\\
210	3.32568302602469\\
225	2.60850759685559\\
240	1.94752940692212\\
255	1.51090187556885\\
270	1.28356049883064\\
285	1.17317158974233\\
300	1.12012438479741\\
315	1.09566163958492\\
330	1.08522120303888\\
345	1.08116239817844\\
360	1.07976014832854\\
375	1.07928363149698\\
390	1.07910121412601\\
405	1.07904119850918\\
420	1.07907939922484\\
435	1.07920710961916\\
450	1.07939854120278\\
465	1.07961558069982\\
480	1.07982166272142\\
495	1.07999080223159\\
510	1.08011056205824\\
525	1.0801803246426\\
540	1.0802074638706\\
555	1.08020335392778\\
570	1.08018014649858\\
585	1.08014838101891\\
600	1.080116144453\\
};
\addlegendentry{$\dot{m}_\mathrm{U}^\mathrm{C}$}

\end{axis}

\end{tikzpicture}%
	\caption{Tracking of a constant reference temperature using DecMPC in a 10-hour simulation: building temperatures and user-requested mass flow rates over time.}
	\label{fig:resultsDecMPCTracking10h}
\end{figure}
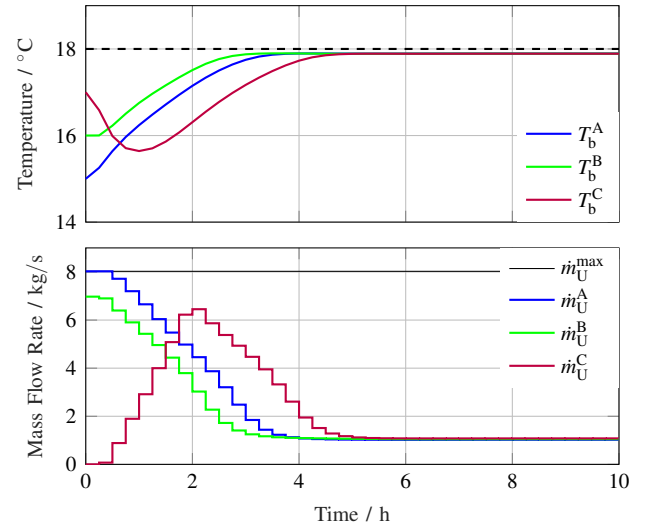

Given the aforementioned operating conditions, it becomes evident that the two control architectures exhibit markedly disparate performance characteristics. In particular, DecMPC operates with an egocentric approach towards its own energy requirements, disregarding the demands and constraints of other buildings in the network. Consequently, when operating with limited resources as in the simulation scenario, a peak demand from buildings $A$ and $B$ results in a lack of mass flow for building $C$, causing its temperature to decrease until the other buildings reduce their energy consumption, as can be observed in \autoref{fig:resultsDecMPCTracking10h}.
The results of the simulation confirm that in this control scheme, a lack of coordination between the controllers has the consequence of rendering the mass flow from the central unit significantly more critical. In the event that the mass flow is insufficient, it will result in the inability to heat the more distant buildings. Conversely, if the energy demand decreases, this will result in a sudden increase in flow through the bypass, which will inevitably lead to substantial energy waste.

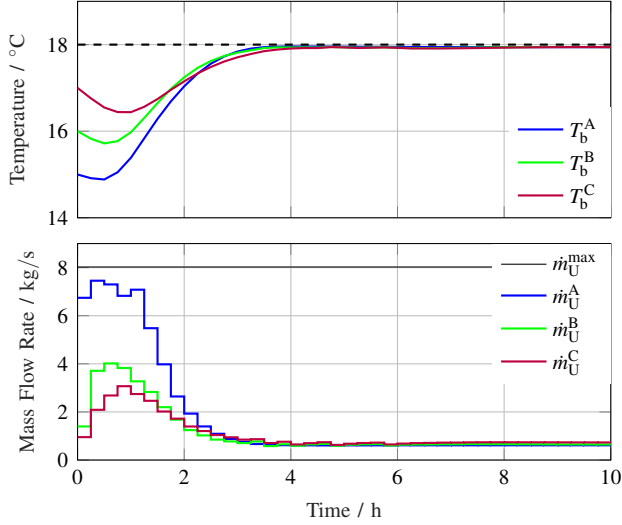
\begin{figure}[H]
	\centering
%
%
\begin{tikzpicture}

\pgfmathsetmacro{\xmin}{0}
\pgfmathsetmacro{\xmax}{600}
\pgfmathsetmacro{\intervalTicks}{2}
\pgfmathsetmacro{\interval}{60*\intervalTicks}
\pgfmathsetmacro{\height}{0.25\textwidth}

\begin{axis}[%
name=temperaturePlot,
width=\linewidth,
height=\height,
unbounded coords=jump,
xmin=\xmin,
xmax=\xmax,
xtick={\xmin,\interval,...,\xmax},
xticklabels=\empty,
ymin=14,
ymax=19,
ylabel style={font=\footnotesize\color{white!15!black}},
ylabel={Temperature / $\si{\celsius}$},
axis background/.style={fill=white},
tick label style={font=\footnotesize},
grid=both,
legend style={at={(1,0)}, anchor=south east, legend cell align=left, draw=none, font=\footnotesize},
]

\addplot [color=blue, thick]
  table[row sep=crcr]{%
0	15\\
15	14.9142115711848\\
30	14.8843511220892\\
45	15.0497208473215\\
60	15.3848074766669\\
75	15.8336039007005\\
90	16.2846515829369\\
105	16.6883144726707\\
120	17.0367119988518\\
135	17.3289466984797\\
150	17.5588273995453\\
165	17.726759137837\\
180	17.8400958525118\\
195	17.907929626799\\
210	17.9414638066719\\
225	17.9529873322952\\
240	17.9573183319649\\
255	17.9568098385762\\
270	17.9556673419756\\
285	17.954351214803\\
300	17.9517914865648\\
315	17.9501912213005\\
330	17.9491222579709\\
345	17.9481499533998\\
360	17.9465144388565\\
375	17.9453241465648\\
390	17.9448502289852\\
405	17.9442724666845\\
420	17.9438217796751\\
435	17.9433069116355\\
450	17.9429561878384\\
465	17.9425438710069\\
480	17.9423038546864\\
495	17.9420521140861\\
510	17.9418965074796\\
525	17.9416854233428\\
540	17.9415125488627\\
555	17.9413264575724\\
570	17.9412031129593\\
585	17.9410434583318\\
600	17.9408954959912\\
};
\addlegendentry{$T_\mathrm{b}^\mathrm{A}$}

\addplot [color=green, thick]
  table[row sep=crcr]{%
0	16\\
15	15.8265020677523\\
30	15.7180270814972\\
45	15.7701816560473\\
60	15.9752230571131\\
75	16.3054631073454\\
90	16.6512705527891\\
105	16.9771580772786\\
120	17.2441340391266\\
135	17.4617478752046\\
150	17.6203628074597\\
165	17.7371601579835\\
180	17.8155720664794\\
195	17.8690256645472\\
210	17.9220189897056\\
225	17.9362301134229\\
240	17.9330930515973\\
255	17.9359703762801\\
270	17.9315807893706\\
285	17.9465121922746\\
300	17.9377386461051\\
315	17.9285597148315\\
330	17.9337070704621\\
345	17.9382309194044\\
360	17.92929617376\\
375	17.9194325654377\\
390	17.9205185972863\\
405	17.9205224446807\\
420	17.9235669899955\\
435	17.9265461178465\\
450	17.9313059947026\\
465	17.9363114494759\\
480	17.939633449632\\
495	17.9425124411567\\
510	17.9443979997653\\
525	17.9455792348122\\
540	17.9459899774659\\
555	17.9457706622168\\
570	17.9449672303683\\
585	17.9436837918481\\
600	17.941986015225\\
};
\addlegendentry{$T_\mathrm{b}^\mathrm{B}$}

\addplot [color=purple, thick]
  table[row sep=crcr]{%
0	17\\
15	16.7596054424757\\
30	16.5481150496825\\
45	16.4419115099126\\
60	16.4394982977427\\
75	16.563119214848\\
90	16.7434173797664\\
105	16.9543256418397\\
120	17.1518010415405\\
135	17.3339238226904\\
150	17.4835226738928\\
165	17.6093581896145\\
180	17.7071996480928\\
195	17.7827360779403\\
210	17.8491513517768\\
225	17.8858227082141\\
240	17.9097367669794\\
255	17.9209501653428\\
270	17.9205812539711\\
285	17.9422230879169\\
300	17.9348286567652\\
315	17.922403052922\\
330	17.928162161558\\
345	17.9346270278448\\
360	17.9252916360274\\
375	17.9120216242371\\
390	17.9115196579266\\
405	17.9111856216104\\
420	17.913766490652\\
435	17.9168232327751\\
450	17.9210832913353\\
465	17.9261246675893\\
480	17.9305297467517\\
495	17.9344010441327\\
510	17.9375281586615\\
525	17.9399324350841\\
540	17.9416423890428\\
555	17.9426908552731\\
570	17.9431124317359\\
585	17.9429436646158\\
600	17.9422264545391\\
};
\addlegendentry{$T_\mathrm{b}^\mathrm{C}$}

\addplot [color=black, dashed, thick]
  table[row sep=crcr]{%
0	18\\
600	18\\
};

\end{axis}

\begin{axis}[%
at={(temperaturePlot.below south)},
anchor=north,
yshift=-0.1cm,
width=\linewidth,
height=\height,
unbounded coords=jump,
xmin=\xmin,
xmax=\xmax,
xtick={\xmin,\interval,...,\xmax},
xticklabels={\xmin,\intervalTicks,...,24},
xlabel style={font=\footnotesize\color{white!15!black}},
xlabel={Time / $\si{\hour}$},
ymin=0,
ymax=9,
ylabel style={font=\footnotesize\color{white!15!black}},
ylabel={Mass Flow Rate / $\si{\kilo\gram/\second}$},
axis background/.style={fill=white},
tick label style={font=\footnotesize},
grid=both,
legend style={at={(1,1)}, anchor=north east, legend cell align=left, draw=none, font=\footnotesize},
]

\addplot [color=black]
  table[row sep=crcr]{%
0	8.02665475463358\\
600	8.02665475463358\\
};
\addlegendentry{$\dot{m}_\mathrm{U}^\mathrm{max}$}

\addplot [const plot, color=blue, thick]
  table[row sep=crcr]{%
0	6.74533272528076\\
15	7.45595277725575\\
30	7.30155809641641\\
45	6.82767604662237\\
60	7.08270437515797\\
75	5.4815830221083\\
90	3.97855999608933\\
105	2.64288020497625\\
120	1.93050881165195\\
135	1.397104747019\\
150	1.09143892815796\\
165	0.875691833147193\\
180	0.758200284325488\\
195	0.666244396429657\\
210	0.631779749694864\\
225	0.618033834015955\\
240	0.609868631183824\\
255	0.611889098381709\\
270	0.608305685486388\\
285	0.606397722739525\\
300	0.610825165802557\\
315	0.609826241805633\\
330	0.611119224770443\\
345	0.607639963401738\\
360	0.612767208479104\\
375	0.611213937860788\\
390	0.613196078091129\\
405	0.611604561581107\\
420	0.613232504979087\\
435	0.612217878657194\\
450	0.613304761254263\\
465	0.612792742726205\\
480	0.61352117156106\\
495	0.61319507108077\\
510	0.613524421423174\\
525	0.61332317873754\\
540	0.613563954623777\\
555	0.613562820351058\\
570	0.613539116459137\\
585	0.613579629601292\\
600	0.613579629601292\\
};
\addlegendentry{$\dot{m}_\mathrm{U}^\mathrm{A}$}

\addplot [const plot, color=green, thick]
  table[row sep=crcr]{%
0	1.39560494842954\\
15	3.71081850658045\\
30	4.01879267181908\\
45	3.82681956447113\\
60	3.27232560906408\\
75	2.8296597618098\\
90	2.1972651452571\\
105	1.67998806002944\\
120	1.24381899972864\\
135	1.02439789535032\\
150	0.85017169436669\\
165	0.775869767838974\\
180	0.705796010273684\\
195	0.772853102518054\\
210	0.581003820697232\\
225	0.658043090488963\\
240	0.594731114409131\\
255	0.650938026416044\\
270	0.669914852667344\\
285	0.596225856212131\\
300	0.638016761751582\\
315	0.654182645380786\\
330	0.660524270874786\\
345	0.60396620248883\\
360	0.638113741982863\\
375	0.645386980474572\\
390	0.650936976780631\\
405	0.657147152332265\\
420	0.656572070853802\\
435	0.665711044957179\\
450	0.661901273090444\\
465	0.660743585927656\\
480	0.658662789982508\\
495	0.656957695817432\\
510	0.65498706620488\\
525	0.653262601570446\\
540	0.65157528751456\\
555	0.65008701665756\\
570	0.648744904941744\\
585	0.647555193233191\\
600	0.647555193233191\\
};
\addlegendentry{$\dot{m}_\mathrm{U}^\mathrm{B}$}

\addplot [const plot, color=purple, thick]
  table[row sep=crcr]{%
0	0.952353774127324\\
15	2.08852894839114\\
30	2.68373466712325\\
45	3.07006428730757\\
60	2.7448348556939\\
75	2.46288414991458\\
90	2.01860552713888\\
105	1.72116060463296\\
120	1.39467296876183\\
135	1.20558691739771\\
150	1.03954201148105\\
165	0.945055811260777\\
180	0.849380146742891\\
195	0.866953993090652\\
210	0.707969016253962\\
225	0.761008276027932\\
240	0.648576040171256\\
255	0.702192092277185\\
270	0.74171396532169\\
285	0.630060560272661\\
300	0.678285176342905\\
315	0.70802083111578\\
330	0.723367772430247\\
345	0.649545882366153\\
360	0.684251077769977\\
375	0.703274118840802\\
390	0.713117276398948\\
405	0.723005726239258\\
420	0.726628057617315\\
435	0.735156977227895\\
450	0.7379285986238\\
465	0.738550738190045\\
480	0.738460084285436\\
495	0.737719558485986\\
510	0.736776641167067\\
525	0.735730813036435\\
540	0.734567823513629\\
555	0.733368580200332\\
570	0.73210219037652\\
585	0.730865682686991\\
600	0.730865682686991\\
};
\addlegendentry{$\dot{m}_\mathrm{U}^\mathrm{C}$}

\end{axis}

\end{tikzpicture}%
	\caption{Tracking of a constant reference temperature using DMPC in a 10-hour simulation: building temperatures and user-requested mass flow rates over time.}
	\label{fig:resultsDMPCTracking10h}
\end{figure}

Conversely, a markedly different behavioral pattern is observed in \autoref{fig:resultsDMPCTracking10h} when DMPC is employed.
In particular, each controller formulates its energy request based not only on its own needs but also on the variables shared with neighboring controllers. This allows controller $A$ to consider the demand from $B$, which in turn is influenced by the requirements of $C$. By penalizing the amount of heat circulating through the bypass in controller $C$ and by adjusting the weights in the cost functions, it is possible to achieve a control action aimed at providing equitable and efficient heating across the entire network, avoiding the situation where a controller close to the heat producer monopolizes the mass flow.

\section{Conclusion \& Outlook}
\label{sec:conclusion}
In this paper, we introduced a distributed nonlinear model predictive control that utilizes a graph-based description of district heating networks. The use of MPC is motivated by the significant flexibility it offers for heating systems, enabling the selection of a cost function that balances multiple objectives. Specifically, our distributed control strategy optimizes heat absorption operations, minimizing discomfort, reducing heat losses, and penalizing auxiliary energy costs for pumping. Another key feature of MPC is its ability to directly incorporate demand forecasts and constraints (such as pressure, mass flow rates, and temperature limitations) into the control actions. \newline
Simulation results show that, with proper tuning, the DMPC approach achieves both good tracking performance and good convergence to consensus. Additionally, with equal energy efficiencies, the DMPC scheme ensures a more equitable distribution of heat. These findings validate the effectiveness of the proposed concept and highlight its potential for improving DHN performance.
Future work may focus on the following areas, identified as potential avenues for further investigation based on the results presented in this paper:
\begin{itemize}
    \item Extend the DMPC framework to include the use of split components introduced in \cite{Model_GB_UsingFlexibility} to describe complex DHNs. Although this extension does not significantly complicate the conceptual framework, it significantly increases the computational costs of the control action because of the much higher number of exchanged variables. Achieving convergence among the distributed controllers in reasonable time should be further investigated in this case.
    \item Integrate the presented DHNs and DMPC into energy hubs frameworks to extend the potential benefits shown in the previous sections to the case of multiple heat-electricity conversion technologies (e.g. heat pumps) and storage units (batteries and thermal storage tanks).
    \item Integrate the dynamic building models to perform thermal peak shaving, thereby reducing the maximum heat supply required during operation. As shown in \cite{MPC_NetworkandBuidling}, it is indeed feasible to leverage the aggregated capacity of buildings to store energy and release it during peaks of demand. 
    \item The ADMM approach used in this paper requires numerous iterations before converging to a neighborhood of the optimal solution, which limits its scalability and applicability for large DHNs. To address this problem, the code can be parallelized, and an accelerated version of the ADMM, as proposed in \cite{SpeedUpMPC_ADMM}, could be implemented to speed up the convergence.
\end{itemize}

\addtolength{\textheight}{-12cm}   


\bibliographystyle{IEEEtran}
\bibliography{Bibliography}

\end{document}